\documentclass{article}
\usepackage{amsmath,float,bm}
\usepackage{hyperref}
\usepackage{authblk,graphicx}

\title{\textbf{Chaos and ergodicity in entangled non-ideal Bohmian qubits}}

\author{A.C. Tzemos \footnote{Corresponding Author: atzemos@academyofathens.gr} and G. Contopoulos}
\affil{Research Center for Astronomy and 
Applied Mathematics of the Academy of 
Athens - Soranou Efessiou 4, GR-11527 Athens, Greece}
\begin{document}
\maketitle



\begin{abstract}
We study the Bohmian dynamics of a large class of bipartite systems of non-ideal qubit systems, by modifying the basic physical parameters of an ideal two-qubit system, made of coherent states of the quantum harmonic 
oscillator. First we study the case of coherent states 
with truncated energy levels and large amplitudes. Then we study 
non-truncated coherent states but with small  amplitudes and finally a combination of the above cases. In all cases we find that the chaotic Bohmian trajectories are approximately ergodic.  We also study the number and the  spatial arrangement of the  nodal points of the wavefunction and their role both  in the formation of 
 chaotic-ergodic trajectories, and in the emergence of ordered trajectories. Our results have strong implications on the dynamical establishment of Born's rule.

\end{abstract}


\vspace{10pt}

\section{INTRODUCTION}

In Bohmian Quantum Mechanics (BQM)  the quantum particles  evolve in spacetime according to the so called Bohmian equations (BEs):
\begin{align}
m_i\frac{dr_i}{dt}=\hbar \Im\left(\frac{\nabla_i\Psi}{\Psi}\right),
\end{align}
where $\Psi$ is the wavefunction of the system under study \cite{Bohm, BohmII} {and} $\Im$ {stands for the imaginary part.} The highly nonlinear character of the BEs allows the coexistence of chaotic and ordered trajectories, which have been studied extensively in the past \cite{iacomelli1996regular, frisk1997properties, falsaperla2003motion, wisniacki2005motion,efthymiopoulos2006chaos, wisniacki2007vortex, efthymiopoulos2007nodal, contopoulos2008ordered, contopoulos2008order, borondo2009dynamical,efth2009,contopoulos2012order}.

The quantum harmonic oscillator (QHO) is the most studied quantum system. In BQM the 2-d and 3-d QHOs have been used extensively in the exploration of  Bohmian chaos, since:
\begin{enumerate}
\item In the case of non-interacting quantum oscillators the corresponding Schr\"{o}\-dinger equation is analytically solvable, i.e. we know with perfect accuracy the guiding wavefunctions of the Bohmian trajectories \cite{ballentine2014quantum}.
\item With a proper choice of their wavefunctions we can observe all possible kinds of Bohmian trajectories: ordered (periodic or not) and chaotic.
\end{enumerate} 
In fact, as it is well known from the very first studies in the field, the nodal points of the wavefunction play a key role for the production of chaotic Bohmian trajectories. According to the `so called nodal point-X-point complex (NPXPC) mechanism' \cite{efth2009,tzemos2018origin}, whenever a quantum particle comes close to a moving nodal point of the wavefunction $\Psi$ it gets scattered by the so called X-point, a stagnant hyperbolic point in the frame of reference of the moving node. Every such scattering process is accompanied by a positive shift of the local Lyapunov characteristic number \cite{Contopoulos200210} and consequently contributes in the production of chaotic Bohmian trajectories. Trajectories that do not approach NPXPCs are ordered.

 
An important {feature} of the non-interacting QHOs is that, in many cases, we can find analytically the positions of the nodal points. This fact  facilitates significantly all the calculations involved in the study of Bohmian chaos in most of our previous works.

A distinguished class of solutions of the quantum harmonic oscillator are the coherent states \cite{garrison2008quantum}. The coherent states are known as the minimum uncertainty states, which exhibit the closest possible behaviour to that of the classical harmonic oscillator. These states are of fundamental importance in Quantum Optics, since they describe under some conditions the state of a laser. Thus it is interesting to study their behaviour  from a Bohmian standpoint.

In a series of previous works \cite{tzemos2019bohmian, tzemos2020ergodicity, tzemos2020chaos, tzemos2021role} we used the coherent states of the QHO  in order to construct qubits \cite{asboth2004coherent} and
 then  studied in detail the Bohmian Dynamics of two such  qubits  for various degrees of quantum entanglement.

The two-qubit system with incommensurable frequencies $\omega_x$ and $\omega_y$ is convenient for both analytical and numerical calculations and has many interesting features:
\begin{enumerate}
\item It has infinitely many nodal points forming straight lines. 
The positions of the nodal points in time can be calculated analytically.
%

\item The infinite number of NPXPCs  was found to be responsible for the production of chaotic-ergodic trajectories, for any non-zero entanglement (for zero entanglement (product states) there are no NPXPCs and all Bohmian trajectories are ordered). In fact we found that for any non-vanishing amount of the entanglement, the chaotic trajectories  are essentially ergodic.
  The proportion of chaotic trajectories increases with entanglement and as we approach the  maximum entanglement all trajectories become chaotic-ergodic. Thus, for strongly entangled states
almost any initial distribution with  $P_0\neq|\Psi_0|^2$  will finally reach the Born Rule (BR) distribution $P=|\Psi|^2$, since in these cases BR is dominated by chaotic-ergodic trajectories.

\item However, when the entanglement is weak the number of ordered trajectories in the Born distribution is significant. Therefore an arbitrary initial distribution will reach the Born distribution only if the ratio between its chaotic and ordered trajectories is approximately the same as that of the Born distribution.


\end{enumerate}

The above results refer to a two-qubit model which consists of perfectly coherent states of two quantum harmonic oscillators in the $x$ and $y$ directions.  But, as it is well known, a coherent state contains an infinite number of energy levels following a Poisson distribution around a mean value \cite{garrison2008quantum}. This infinite number of energies is responsible for the  infinite number of the nodal points in this model, something crucial for the quick emergence of the ergodicity of the chaotic trajectories. 

It is natural now to ask how general are our results i.e.
 what happens if we  work with wavefunctions that deviate from a two-qubit model.
 
  In the present paper we study the dynamics of generic bipartite Bohmian systems of quantum oscillators, by considering two basic modifications of the two-qubit model: a)  coherent states with truncated energy levels and b) qubits with overlapping basis states.

%

The structure of the paper is the following: In section 2
we present the general form of the bipartite system of quantum harmonic oscillators and point out the key physical parameters. In section 3  we consider states with truncated  energy levels in the two-qubit model and  clarify the contribution of the various energy levels in the long time behaviour of the trajectories, for different values of the entanglement. Then in section 4 we compare our results with those of more general cases that deviate from a two-qubit system, namely:  coherent states with common small amplitudes in $x$ and $y$ and  coherent states with different amplitudes along the $x$ and $y$ axes. In section 5 we consider truncated coherent states with small amplitudes and finally,   in section 6 we draw our  conclusions.

\section{A BIPARTITE SYSTEM OF QUANTUM HARMONIC OSCILLATORS}

The coherent states of the 1-d quantum harmonic oscillator, corresponding to the classical system \begin{align}\label{H1}
H=\frac{1}{2}m\omega^2x^2+\frac{p^2}{2m},
\end{align}
 are
defined as the eigenstates of the anihillation operator $\hat{a}$: 
\begin{align}
\hat{a}|\alpha\rangle=A|\alpha\rangle,
\end{align}
where the eigenvalue $A$ is, in general, a complex number since $\hat{a}$ is not hermitian, i.e. $A=|A|\exp(i\theta)$, where $|A|$ is the amplitude and $\theta$ the phase of the state $|\alpha\rangle$. The coherent states are written in the basis of Fock states as:
\begin{align}\label{nrep}
|\alpha\rangle=e^{-\frac{1}{2}|A|^2}\sum_{n=0}^{\infty}\frac{A^n}{\sqrt{n!}}|n\rangle,
\end{align}
where $|n\rangle$ are the eigenvectors of the Hamiltonian  operator $\hat{H}=\hbar\omega(\hat{a}^{\dagger}\hat{a}+\frac{1}{2})$.
The infinitely many energy eigenstates inside a coherent state follow Poissonian statistics \cite{garrison2008quantum}, i.e. the probability {of detecting energy level} $n$ in the state $|\alpha\rangle $ is 
\begin{align}
P(n)=|\langle n|\alpha\rangle|^2=\frac{e^{-\langle n\rangle}\langle n\rangle ^n}{n!},
\end{align} The mean value $\langle n\rangle$ and the variance $(\Delta n)^2$  are both constant, equal to $|A|^2$, and  $\sum_{n=0}^{n=\infty}P(n)=1$  (see Fig.~\ref{Poisson}). By taking the inner product $\langle x|\alpha\rangle$ and using the fact that the energies of the quantum harmonic oscillator are given by $E_n=\hbar\omega(n+\frac{1}{2}), n=0, 1, 2,\dots$,
one finds that the time dependent wavefunction of an 1-d coherent state is given by:
\begin{align}\label{coh}
Y(x,t)=e^{-\frac{1}{2}a_0^2}e^{\frac{-i\omega_x t}{2}}\sum_{n=0}^{n_f}\frac{(a_0e^{i\sigma_x} e^{-i\omega_x t})^n}{\sqrt{n!}}\psi_n(x),
\end{align}
where 
\begin{align}
\psi_n(x)=\frac{1}{\sqrt{2^nn!}}\left(\frac{m_x\omega_x}{\pi\hbar}\right)^{\frac{1}{4}}e^{-\frac{m_x\omega_x x^2}{2\hbar}}H_n\left(\sqrt{\frac{m_x\omega_ x}{\hbar}}x\right),n=0, 1, 2,\dots,
\end{align}
{and}
$H_n(k)=(-1)^ne^{k^2}\frac{d^n}{dk^n}\left(e^{-k^2}\right)$ {are the corresponding Hermite polynomials}.
{In the ideal coherent state we have} $n_f=\infty$, while
 $a_0=|A(0)|,  \sigma_x, \omega_x$ {are the initial values of the amplitude, the initial phase, and the frequency of the oscillator.}
{Similarly there is an expression for} ${Y(y,t)}$ with $b_0, \sigma_y, \omega_y$ instead of $a_0, \sigma_x$ and $\omega_x$.
 In our calculations we take $m_x=m_y=\hbar=1$.

In our previous papers \cite{tzemos2019bohmian,tzemos2020chaos,
tzemos2020ergodicity,tzemos2021role}, we studied  the Bohmian Dynamics of entangled coherent states of the form
\begin{align}\label{wavefunction}
\Psi=c_1Y_R(x,t)Y_L(y,t)+c_2Y_L(x,t)Y_R(y,t),
\end{align}
where 
\begin{align}
Y_R(x,t)=Y(x,t;\sigma_x=0), 
Y_L(x,t)=Y(x,t;\sigma_x=\pi)
\end{align}
and similarly for the coordinate $y$.
In the limit of infinite energies $n_f=\infty$  and with ${a_0=b_0}$  { sufficiently large},  $Y_R$ and $Y_L$ define the orthogonal basis states of a qubit, where $Y_R$ (or $Y_L$) refers to a Gaussian blob of a coherent state starting at $t=0$ on the right (or on the left) of the center of the oscillation. Consequently, with a proper choice of the coefficients $c_1,c_2$ for which we have $|c_1|^2+|c_2|^2=1$, the wavefunction $\Psi$ describes a two-qubit model covering all the possible degrees of entanglement. We chose to work with real $c_1, c_2$ and used $c_2$ as the entanglement control parameter. For $c_2=0$ the entanglement is zero (product state) and for $c_2=\sqrt{2}/2$ the entanglement is maximum (Bell state).\footnote{In  \cite{tzemos2019bohmian} we also considered  wavefunctions of the form  $\Phi=c_1Y_R(x,t)Y_R(y,t)+c_2Y_L(x,t)Y_L(y,t)$ but the results are qualitatively very similar to those of the case of Eq.~\ref{wavefunction}.} 

The key physical parameters for the qubit character of this model are a) the amplitudes of the oscillators: the larger the amplitudes the smaller the overlap between the two basis states of the qubits. b) the energy levels of the coherent state, which in the nontruncated case they are infinite.  Thus, we first  work with common large  amplitudes $a_0=b_0=2.5$  for both oscillators  and truncate the energy levels $n_f$ (section 3). Then we work with non-truncated states but with small amplitudes (section 4) and finally with truncated states and small amplitudes (section 5).

\section{TRUNCATED COHERENT STATES}

\begin{figure}[!h]
\centering
\includegraphics[scale=0.25]{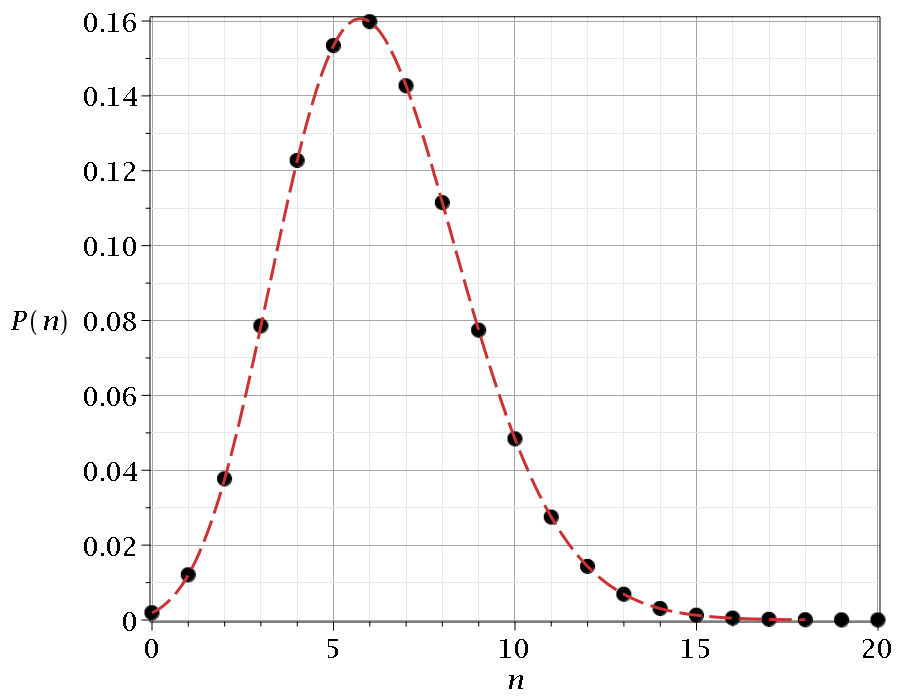}
\caption{The energy level distribution $P(n)$ of a 1-d coherent state with $a_0=2.5$. In this case the average value of $n$ is $\langle n\rangle=6.25$.}\label{Poisson}
\end{figure}

The energy levels inside the coherent states of every qubit  follow a Poissonian distribution as the one shown in Fig.~\ref{Poisson}. 

We now study  the case of truncated coherent states, where $n_f$ is a finite positive integer. The non-classical properties of the truncated coherent states and their importance in Quantum Optics have already been studied in \cite{chung2014even, sivakumar2014truncated, chung2020two, chung2020truncated}. Here we study the dynamics of the corresponding Bohmian trajectories. In these states the spectrum of the energies is smaller than the Poissonian one of Fig.~\ref{Poisson}. In particular for $n_f=2$ we cover only $5.1\%$ of the total distribution, for $n_f=4$ we cover  $25.3\%$ etc. In the maximum value  that we consider in this paper, $n_f=12$, we  cover $98.7\%$ of the full energy distribution and practically recover the full coherent state system.

\begin{figure}[H]
\centering
\includegraphics[scale=0.2]{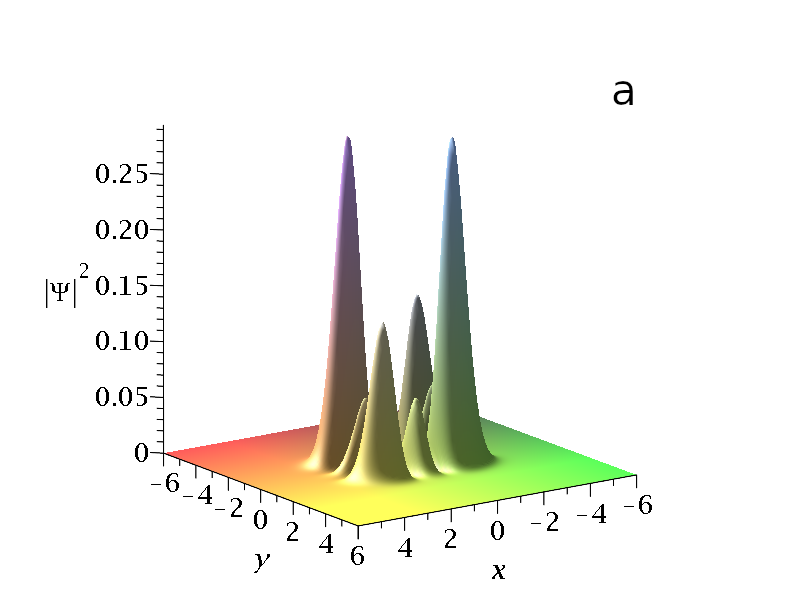}
\includegraphics[scale=0.2]{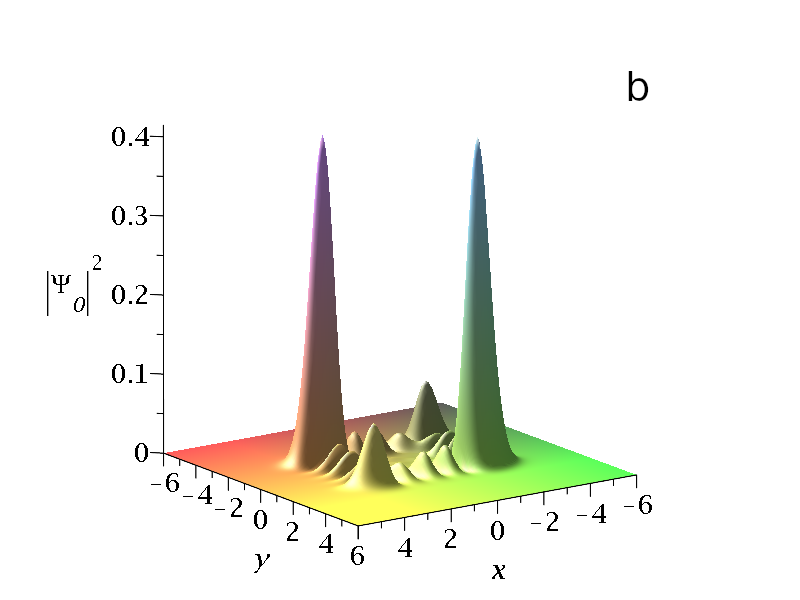}\\
\includegraphics[scale=0.2]{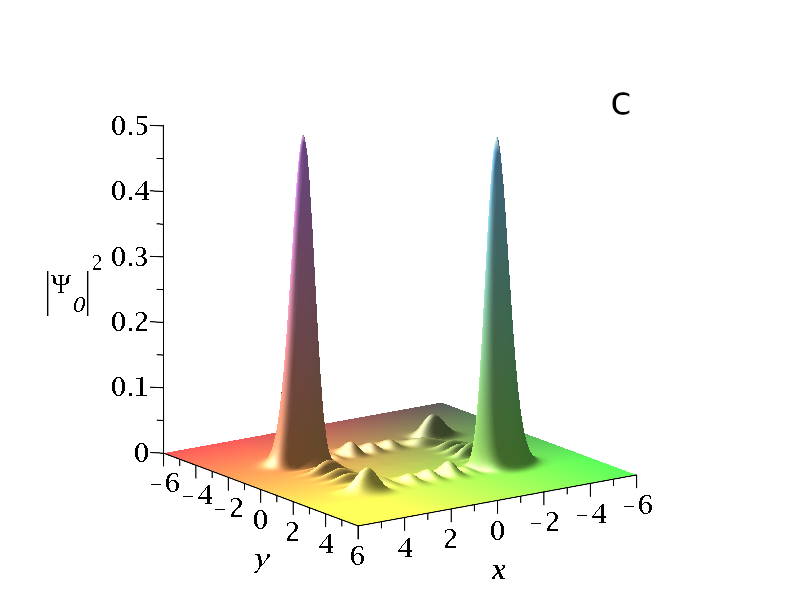}
\includegraphics[scale=0.2]{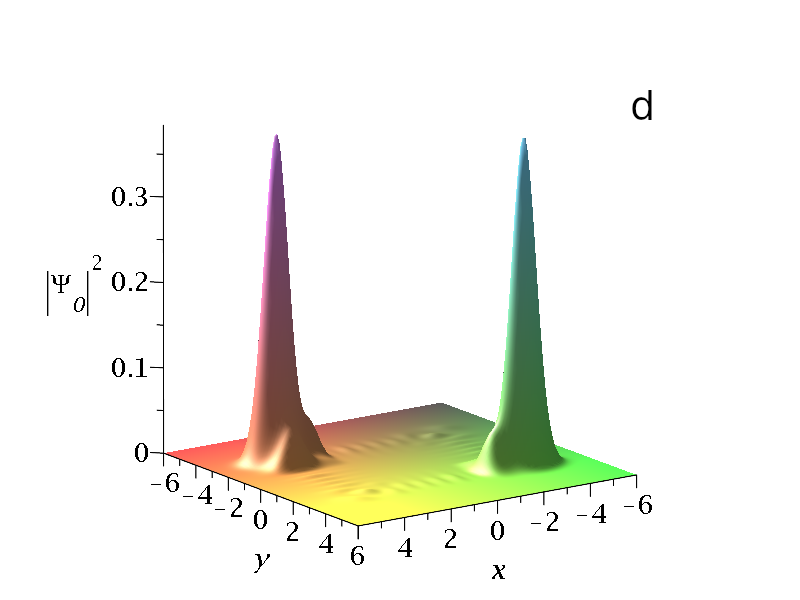}

\caption{The initial probability density $P_0(x,y)=|\Psi_0|^2$ of the maximally entangled two-qubit state  ($c_2=\sqrt{2}/2$) with $a_0=2.5,\omega_x=1,\omega_y=\sqrt{3}$, in truncated coherent states with a) $n_f=2$, b) $n_f=4$, c) $n_f=6$ and d) $n_f=10$.  For $n_f\geq 12$ we recover the results of the full coherent state qubits (Fig.~1 of \cite{tzemos2020chaos}).}\label{Psitet}
\end{figure}
 First we calculate the normalized initial probability density $|\Psi_0|^2$ of finding a quantum particle at a certain point of the configuration space, which represents  an initial particle distribution following the Born rule $P_0=|\Psi_0|^2$. In Fig.~\ref{Psitet} we show the case of the maximally  entangled state ($c_2=\sqrt{2}/2$) for various truncations.  
We observe two equal  main blobs symmetric with respect to the origin. However, while for $n_f=\infty$ we have  only these two blobs, when $n_f$ is small we have a number of secondary blobs, which are relatively large for $n_f=2$ and they become smaller as $n_f$ increases. For $n_f> 10$ they practically disappear and we recover the probability density of the non-truncated case.

{Furthermore we have nodal points when $\Psi_{Re}=\Psi_{Im}=0$, whose number increases as  $n_f$ increases and tends to infinity as $n_f\to \infty$ and whose positions are  aligned on a straight line when $n_f= \infty$ \cite{tzemos2021role}.}

  
\begin{figure}[H]
\centering
\includegraphics[scale=0.2]{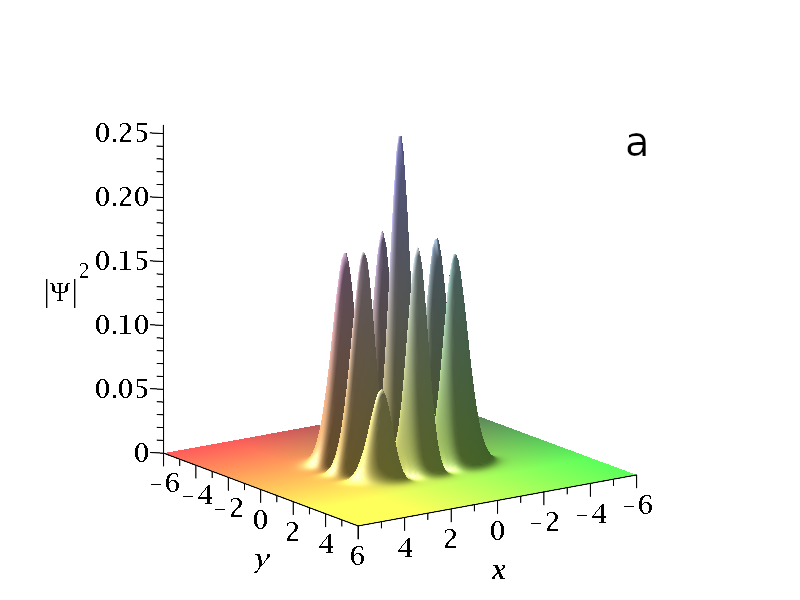}
\includegraphics[scale=0.2]{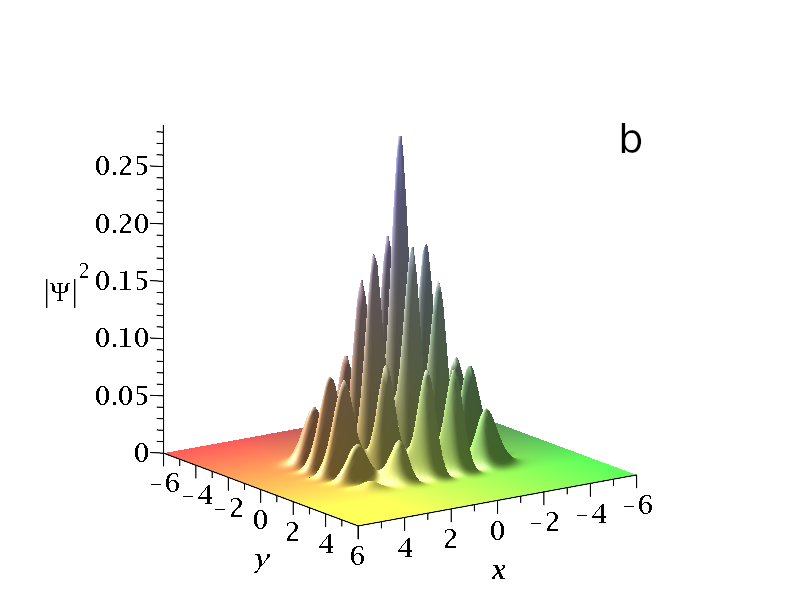}\\
\includegraphics[scale=0.2]{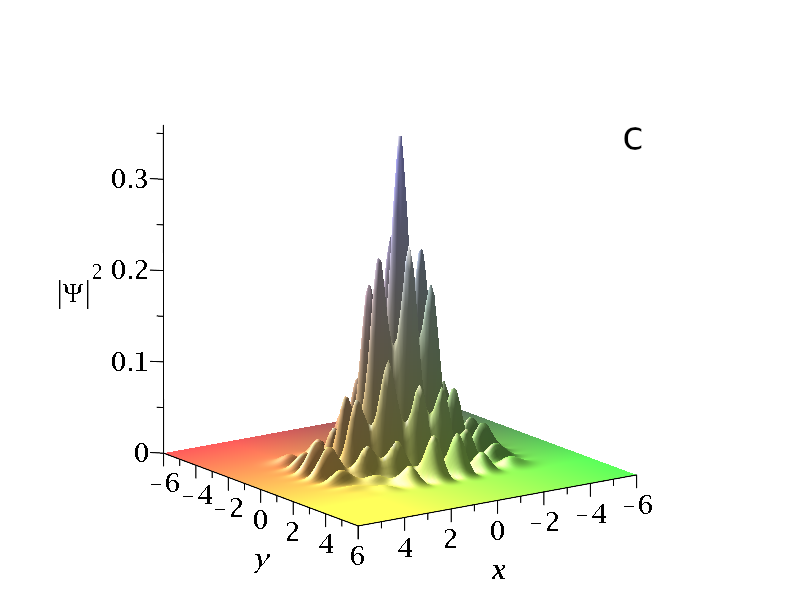}
\includegraphics[scale=0.2]{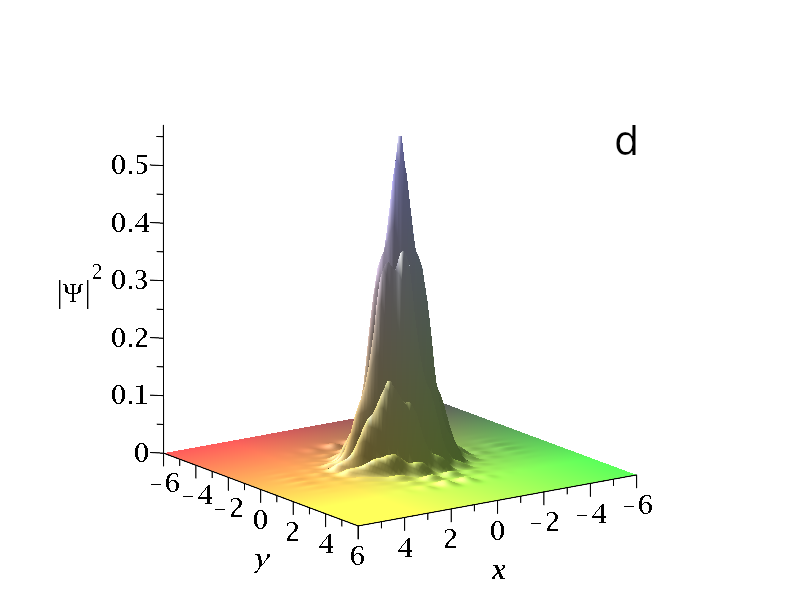}

\caption{The probability density $P(x,y)=|\Psi|^2$ of the maximally entangled two-qubit state  ($c_2=\sqrt{2}/2$) with $a_0=2.5,\omega_x=1,\omega_y=\sqrt{3}$ at $t=4.58$, when the two main blobs collide, in truncated coherent states with a) $n_f=2$, b) $n_f=4$, c) $n_f=6$ and d) $n_f=10$.}\label{krousi}
\end{figure}

At $t=0$ the two blobs are at their maximum distance from the origin and, as the time increases, they move around the origin following approximately Lissajous figures (see Eq.~\ref{Lf})\footnote{{We note that in the case of product states the Bohmian trajectories form exact Lissajous curves \cite{tzemos2019bohmian, tzemos2020ergodicity}.}}. The support of the wavefunction, i.e. the region where $|\Psi|^2$ is relatively large, is approximately a parallelogram around the origin (our plots  {cover} the regions with $|\Psi|^2\geq 10^{-5}$). At particular times the blobs collide close to the origin (Fig.~\ref{krousi}). Such a collision appears for the first time at $t\simeq 4.58$ in all cases \cite{tzemos2020ergodicity, tzemos2020chaos} and they form a complicated pattern of secondary blobs for a small time interval. Between these blobs there are nodal points that scatter the particles of the blobs, so that all their trajectories become chaotic. After the collision  the  blobs are formed again and collide from time to time.

\begin{figure}[H]
\centering
\includegraphics[scale=0.2]{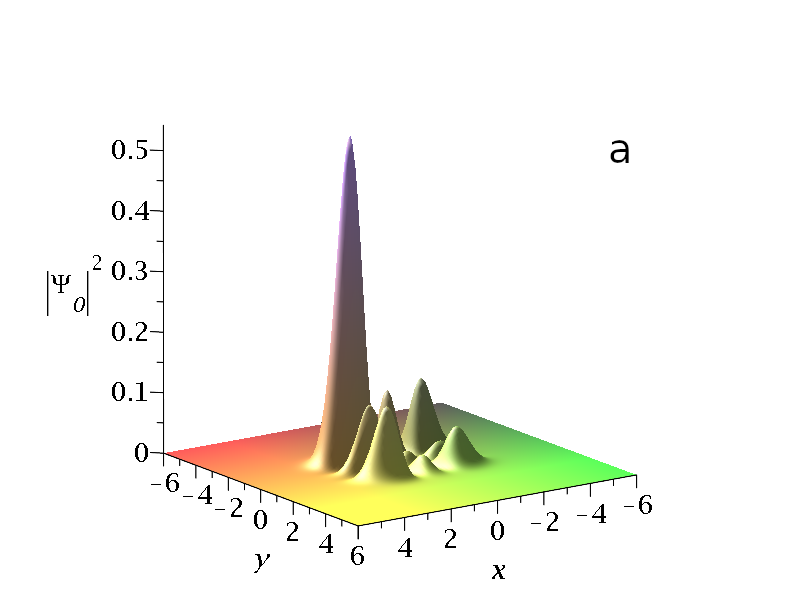}
\includegraphics[scale=0.2]{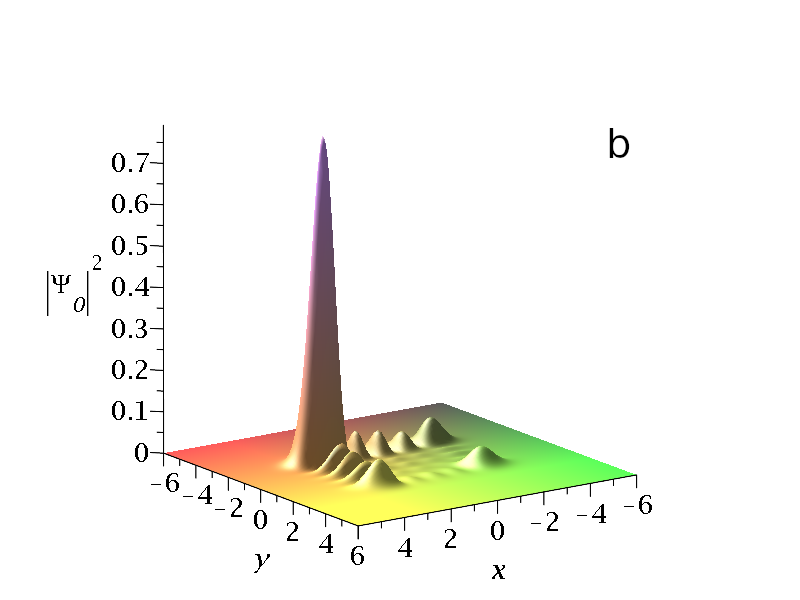}\\
\includegraphics[scale=0.2]{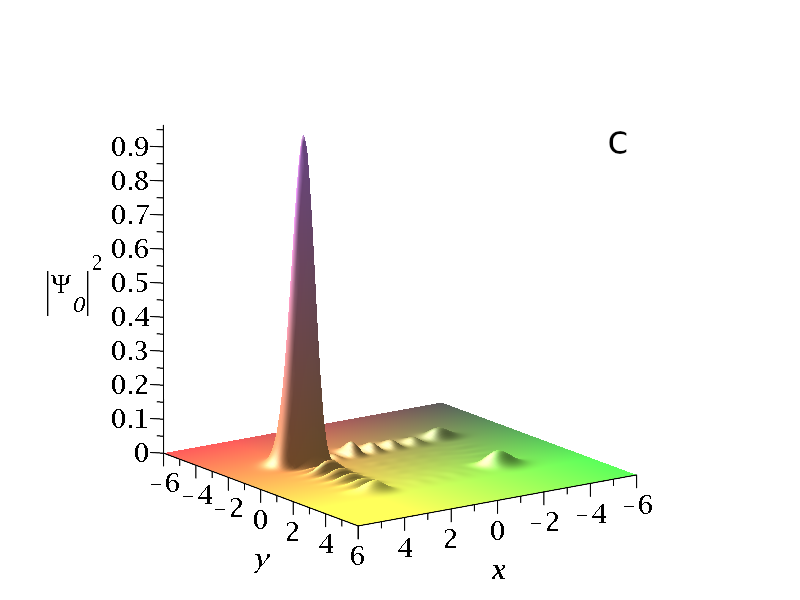}
\includegraphics[scale=0.2]{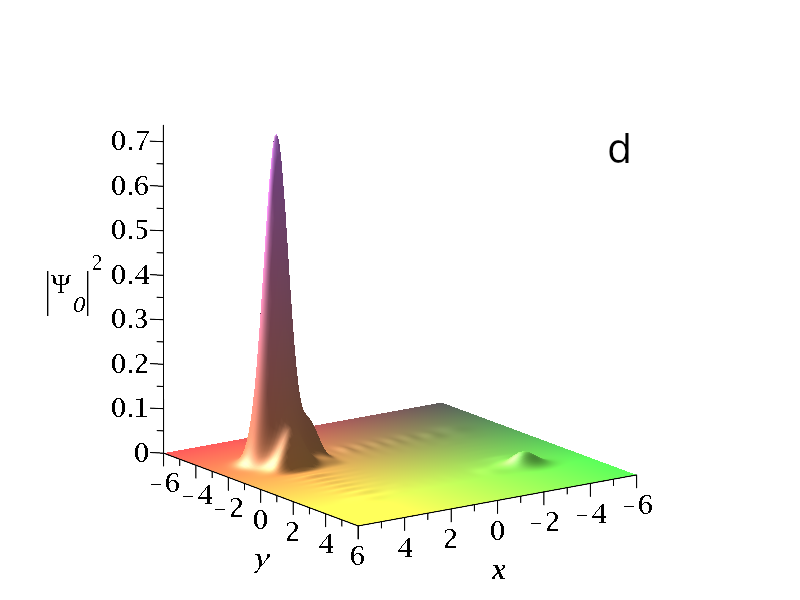}

\caption{The initial probability density $P_0(x,y)=|\Psi_0|^2$ of a weakly entangled two-qubit state  ($c_2=0.2$) with $a_0=2.5,\omega_x=1,\omega_y=\sqrt{3}$,  in truncated coherent states with (a) $n_f=2$, (b) $n_f=4$, (c) $n_f=6$ and (d) $n_f=10$. }\label{Psitet_weak}
\end{figure}

On the other hand, we know from our previous works (e.g. \cite{tzemos2021role}) that when $c_2$ is not maximum , the two main blobs are not equal. We found that the same holds for the small secondary blobs of the truncated cases: as $c_2$ decreases all initial blobs (main and secondary) of the fourth quadrant ($x>0,y<0$) become larger and those of the second quadrant ($x<0,y>0$) become smaller, as in the non-truncated system. E.g.  the  form of  $P_0$  for  $c_2=0.2$ and for  various truncations is shown in Fig.~\ref{Psitet_weak}.

\begin{figure}[H]
\centering
\hspace{-2cm}\includegraphics[scale=0.18]{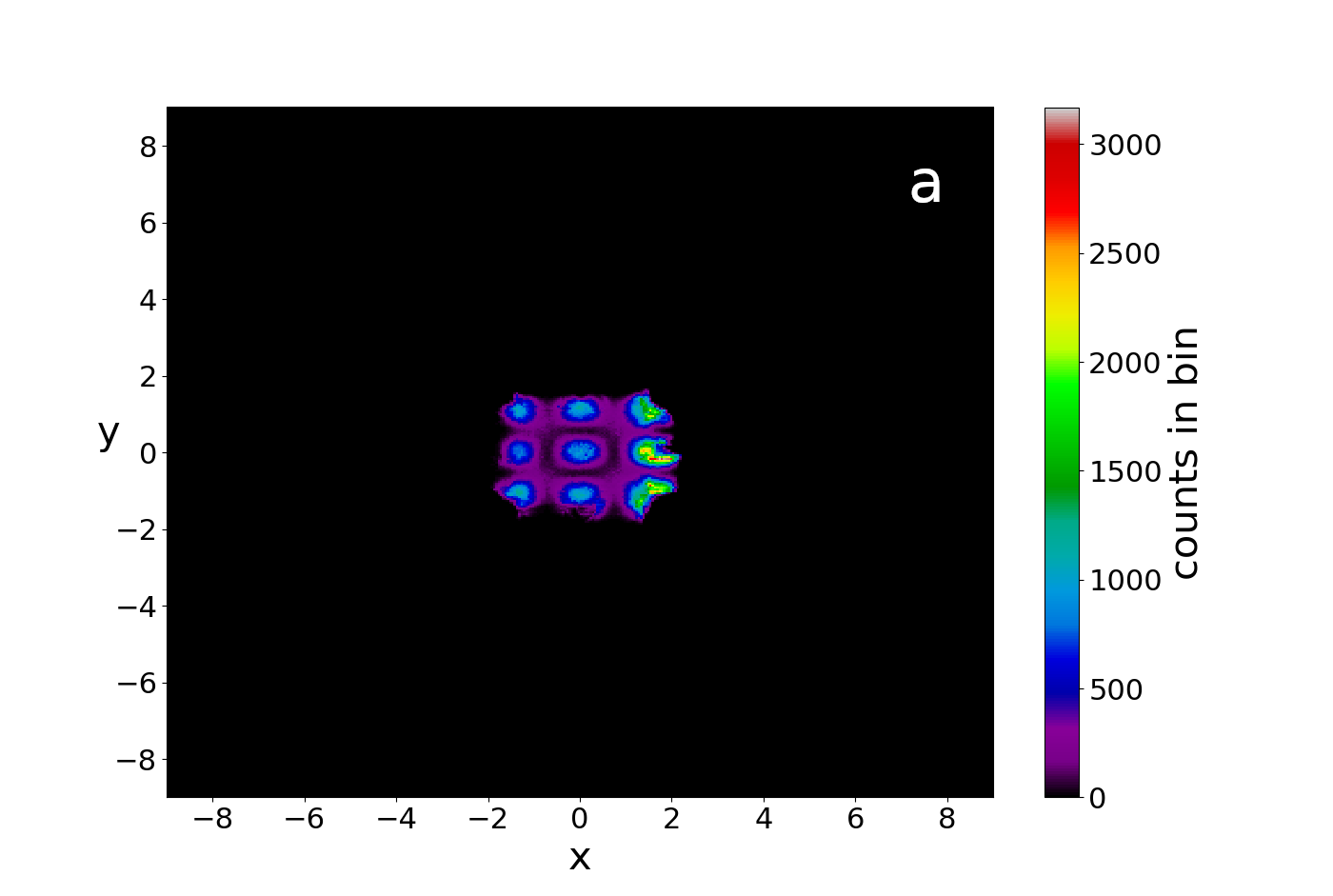}
\includegraphics[scale=0.18]{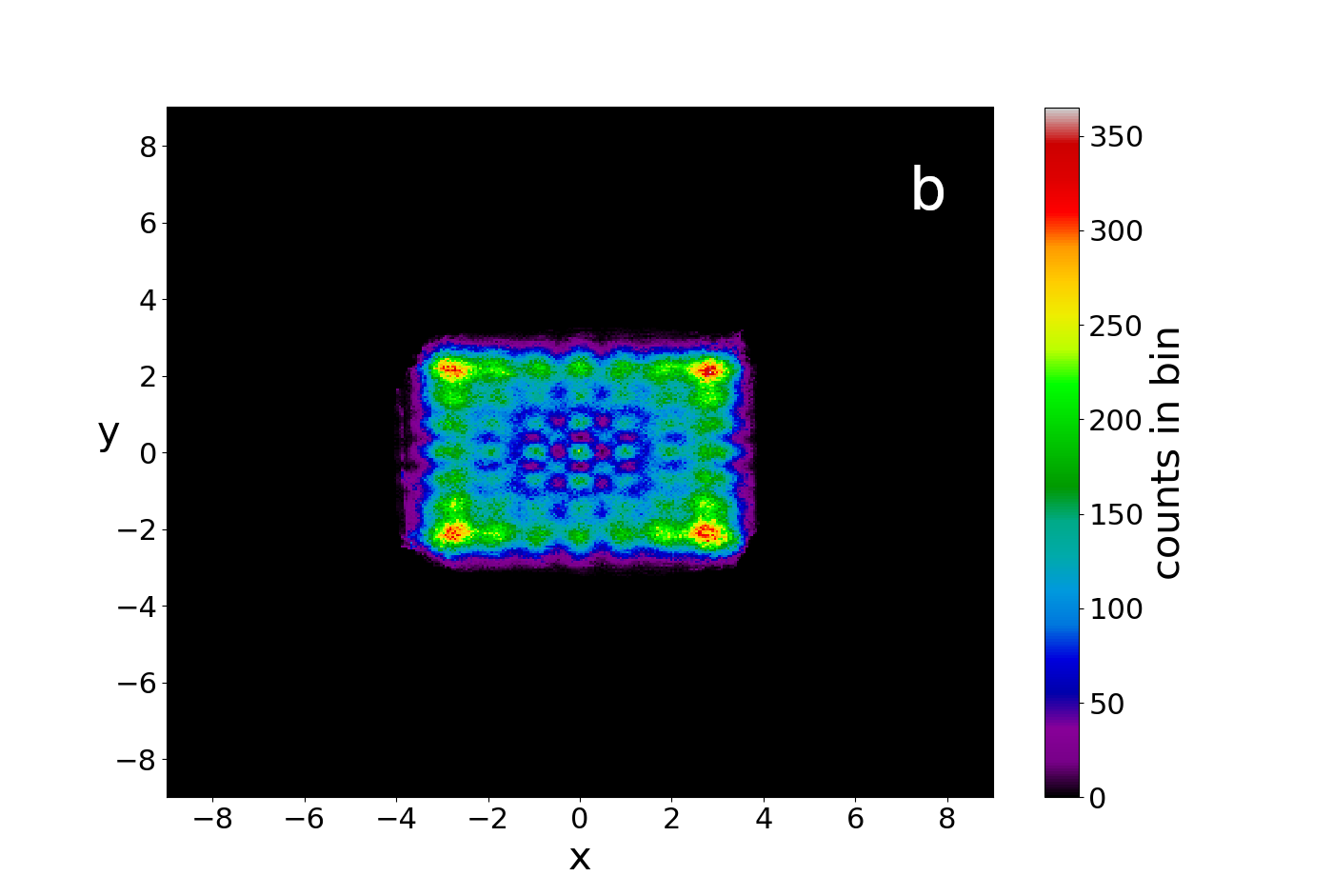}\\
\hspace{-2cm}\includegraphics[scale=0.18]{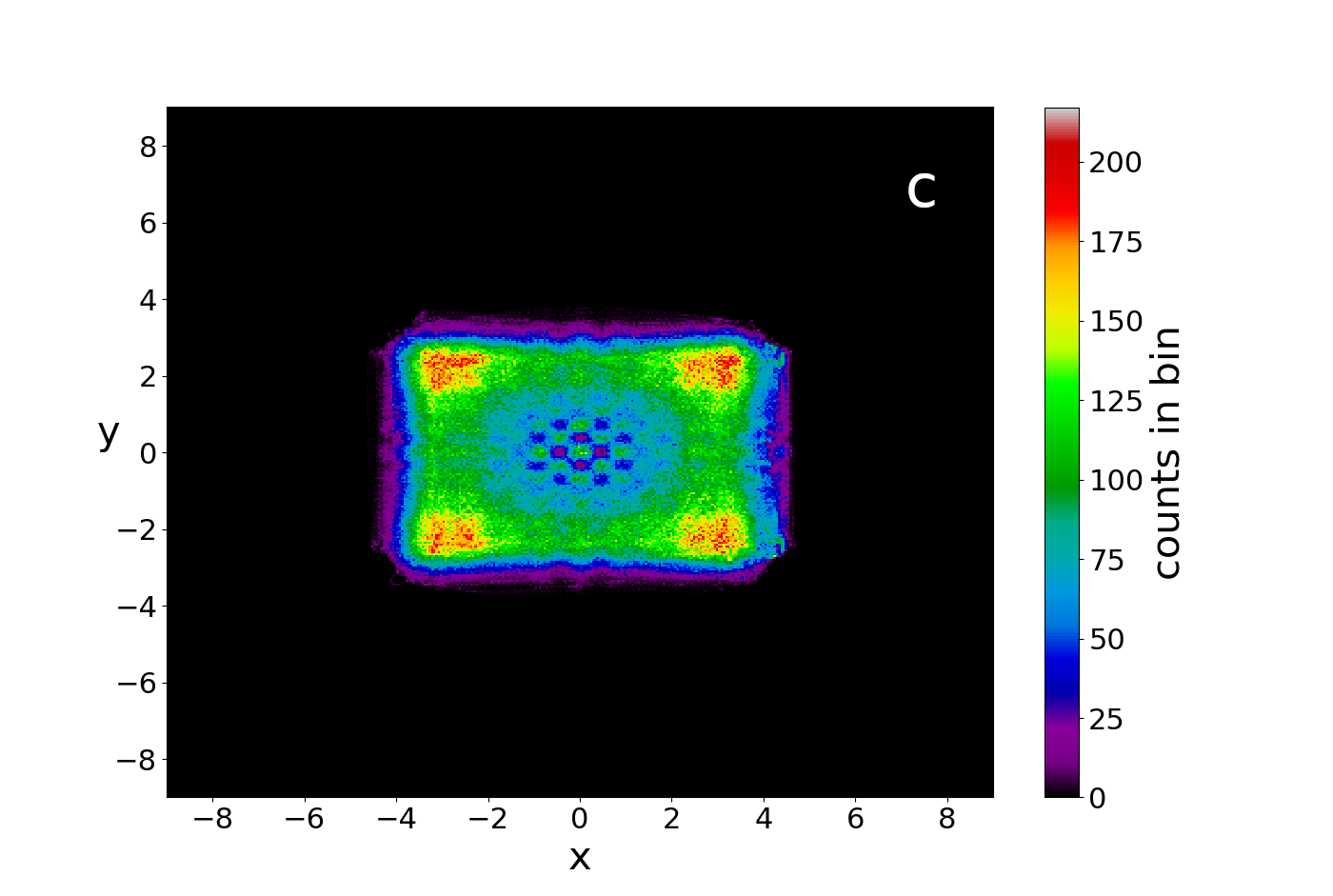}
\includegraphics[scale=0.18]{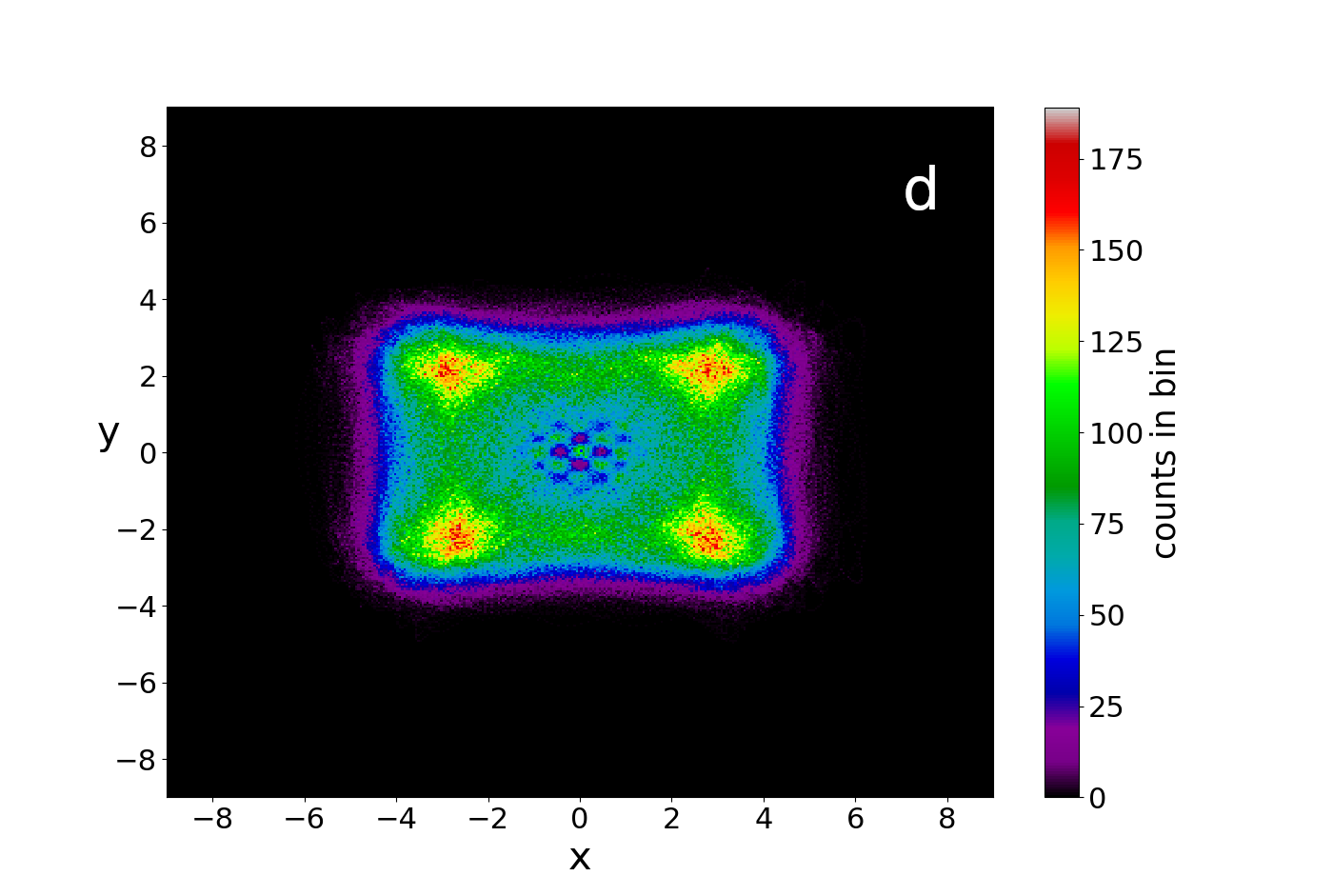}
\caption{Maximally entangled state for (a)  $n_f=2$, (b) $n_f=6$, (c) $n_f=8$ and (d) $n_f=\infty$. We observe how the support of the wavefunction and consequently the range of motion, increases as $n_f\to\infty$.}\label{max}
\end{figure}

Then we check the patterns of the limiting distributions of the points of the  chaotic trajectories in various cases of the truncated states, using the methods developed in our previous papers \cite{tzemos2020ergodicity,tzemos2020chaos}.
Namely, we make colorplots which depict how many times a trajectory has passed through the cells of a $360\times 360$ square grid for $x,y \in[-9,9]$ that covers the support of the wavefunction and compare the underlying arrays of the colorplots.

First we study the maximally entangled state ($c_2=\sqrt{2}/2$). We compute the distribution  of the points of a chaotic trajectory, taken at every $\Delta t=0.05$ for times up to $t=10^5$ and compare their pattern (Figs.~\ref{max}a,b,c) with the corresponding distribution of  the non-truncated system (Fig.~\ref{max}d). We find that the distribution for $n_f=2$ (Fig.~\ref{max}a) is quite different from that for $n_f=\infty$  (Fig.~\ref{max}d) and has a significantly reduced size in the configuration space, which corresponds to the size of the support of the truncated wavefunction\footnote{In this model the chaotic trajectories cover in the long run all the support of the wavefunction. Therefore, the size of the distribution of points of a chaotic trajectory is  essentially the same to the size of the support of $\Psi$.}.  In Fig.~\ref{max}b we see that for $n_f=6$ the size of the distribution is  closer to that of the non-truncated state:  their central parts, close to the origin look very similar and there exist four regions of high concentration close to the corners of the distribution. Finally, in Fig.~\ref{max}c we see that for $n_f=8$  we approach closely the $n_f=\infty$ distribution.

\begin{figure}[H]
\centering
\includegraphics[scale=0.3]{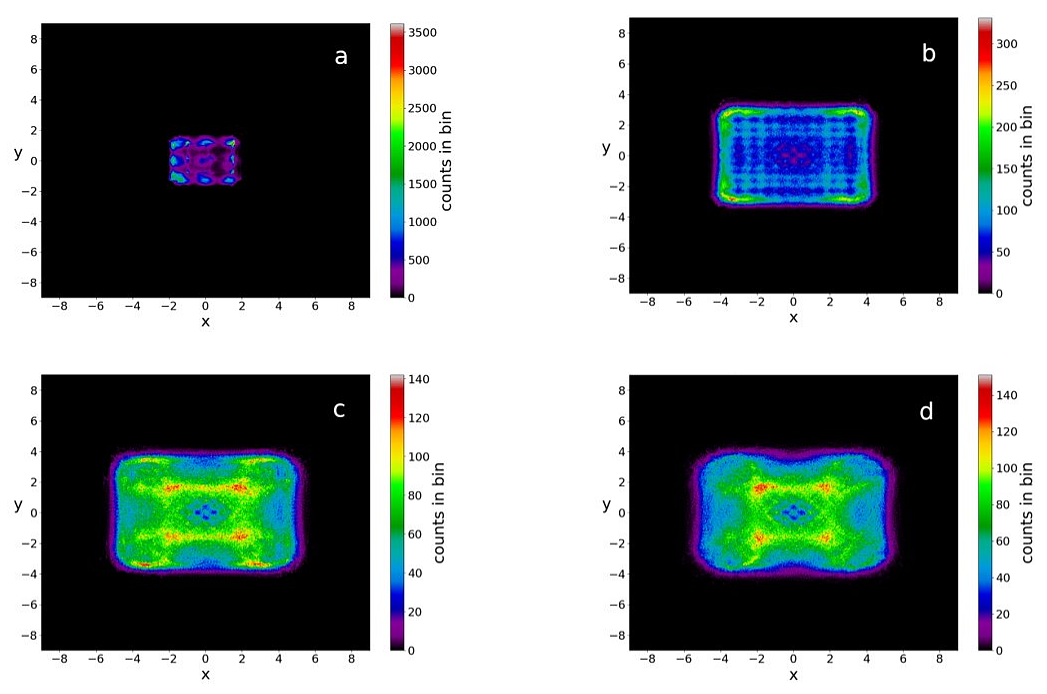}
\caption{The trajectory with $x_0=0.1, y_0=0.4$ of the weakly entangled state $c_2=0.2$ up to $t=10^5$ in the cases a) $n_f= 2$, b) $n_f= 8$, c) $n_f=12$ and d) $n_f=\infty$.}\label{weak}
\end{figure}

Similar results appear in partially entangled states. E.g. the distribution of the points of a trajectory in the case $c_2=0.2$ approaches  the coherent state distribution  as $n_f$ increases (Fig.~\ref{weak}). However, in this case the pattern of the final distribution of chaotic trajectories for $n_f=\infty$ (Fig.~\ref{weak}d) is different from that of Fig.~\ref{max}d (smaller distances of the 4 regions of maximum concentration), as it was already  shown  in \cite{tzemos2020chaos}.

\begin{figure}[H]
\centering
\includegraphics[scale=0.35]{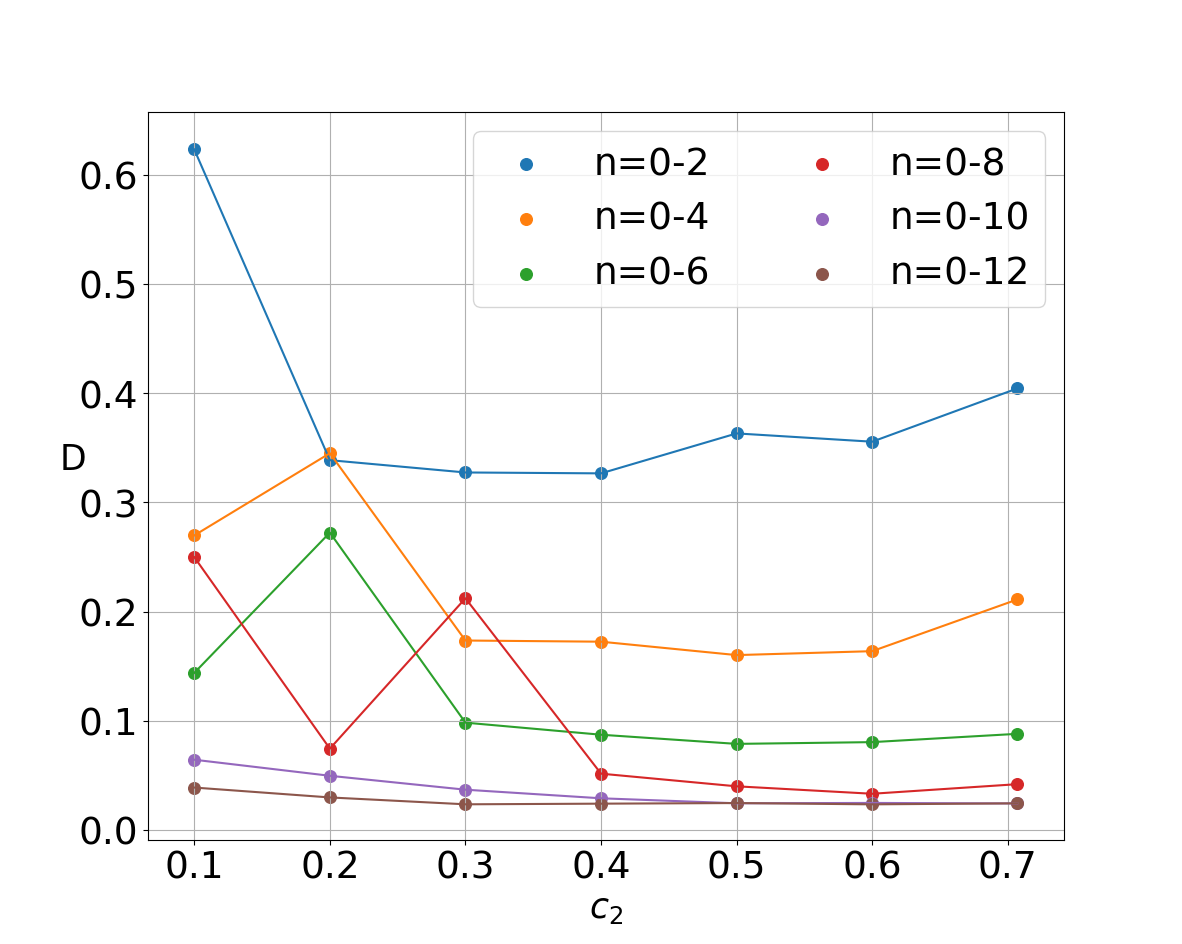}
\caption{Frobenius distance between the final point distribution ($t=10^5$) of a chaotic trajectory with initial conditions ($x_0=0.1, y_0=0.4$), {for several ranges of energy levels  (starting from $n=0$) in the truncated coherent states} and for given entanglements  ($c_2$), from the corresponding complete coherent state.}\label{sigrisi0}
\end{figure}

These differences are established numerically by calculating the Frobenius norm \cite{tzemos2020chaos, tzemos2021role,strang1993introduction} between the underlying matrices of the various color plots\footnote{The Frobenius norm represents the distance between two matrices $A, B$ and is equal to $||A-B||=\sqrt{tr[(A-B)^{\dagger}(A-B)]}$. We note that, in order to compare distributions by the same number of points we always divide the number of counts in the bins with the integration time.}. A detailed calculation is shown in Fig.~\ref{sigrisi0}, where we plot the distance of the pattern established after a time $t=10^5$ between chaotic trajectories in the full coherent state and the truncated coherent states with $n_f=2,4,6,8,10$ and $12$,  for various degrees of the entanglement.

We see that,  for a given value of $n_f$, the values of $D$ decrease in general, or remain roughly constant as the entanglement increases beyond the value $c_2=0.3$.  On the other hand, for any given $c_2$, $D$ tends to zero as $n_f$ increases. Thus we note that the main differences between the various patterns of points are due to the truncation order $n_f$ and not to the entanglement.
Namely the values of $D$ for $n_f\geq 10$ become smaller than $D=0.05$ for all the values of $c_2$, and $D$ becomes even smaller for larger $n_f$.

{Furthermore, for every value of $n_f$ and $c_2$, we  find  that the Frobenius distance  of two different chaotic trajectories is, in general, less than $D=0.015$. Namely, the two trajectories reach practically the same limiting distribution of points. }

{In the case of maximum entanglement ($c_2=\sqrt{2}/2$) all the trajectories inside the support of the wavefunction are chaotic and ergodic. Consequently, any arbitrary initial distribution of particles will reach in the long run the BR distribution, as in the case of the full coherent state system.} However we note that there is a difference: in the non-truncated system we have infinite nodal points outside the support of the wavefunction, while for a small $n_f$ there are only a few nodal points for every time $t$. Thus in the course of time there are large zones in the configuration space that do not contain nodal points (as we will see in Fig.~\ref{komboi_ap}a of section 5).  This implies the possibility of the  existence of ordered trajectories inside them. As $n_f$ increases these zones are gradually  filled with nodal points and in the limit $n_f\to\infty$ there are no ordered trajectories outside the support of the wavefunction. Consequently, in the maximally entangled state of truncated states with small $n_f$ we have ordered trajectories but these are outside the support of the wavefunction, therefore they do not contribute significantly in the Born distribution. Thus Born's rule is accessible in the long run  by all arbitrary initial distributions inside the support of the wavefunction, which is smaller than that of the non-truncated system.

On the other hand, for small $c_2$ we have ordered trajectories inside the support of the wavefunction, as in the non-truncated case, due to the different size of the blobs. Namely, the trajectories that start close to the top of the leading blob do not undergo scattering processes during the collisions of the blobs. This is due to the fact that the higher levels of the leading blob are practically unaffected by the collisions. Therefore for weakly entangled states Born's rule is accessible only by initial distributions with the same ratio between chaotic and ordered trajectories with the BR distribution, as in the non-truncated system.

 The truncated coherent states deviate from Born's rule in the same way as in the case of the full coherent states considered in \cite{tzemos2020chaos}\cite{tzemos2020ergodicity}. Namely, the chaotic trajectories of the truncated states for any value of the entanglement parameter $c_2$ are ergodic, but the patterns of their points are different from the patterns of the trajectorires of the full coherent state, This difference decreases as $n_f$ increases and tends to zero as $n_f$ tends to infinity.

We have made similar calculations by truncating the coherent states not only from above (below $n_f=\infty$), but also from below (near $n_f=0$). 

\begin{figure}[H]
\centering
\includegraphics[scale=0.35]{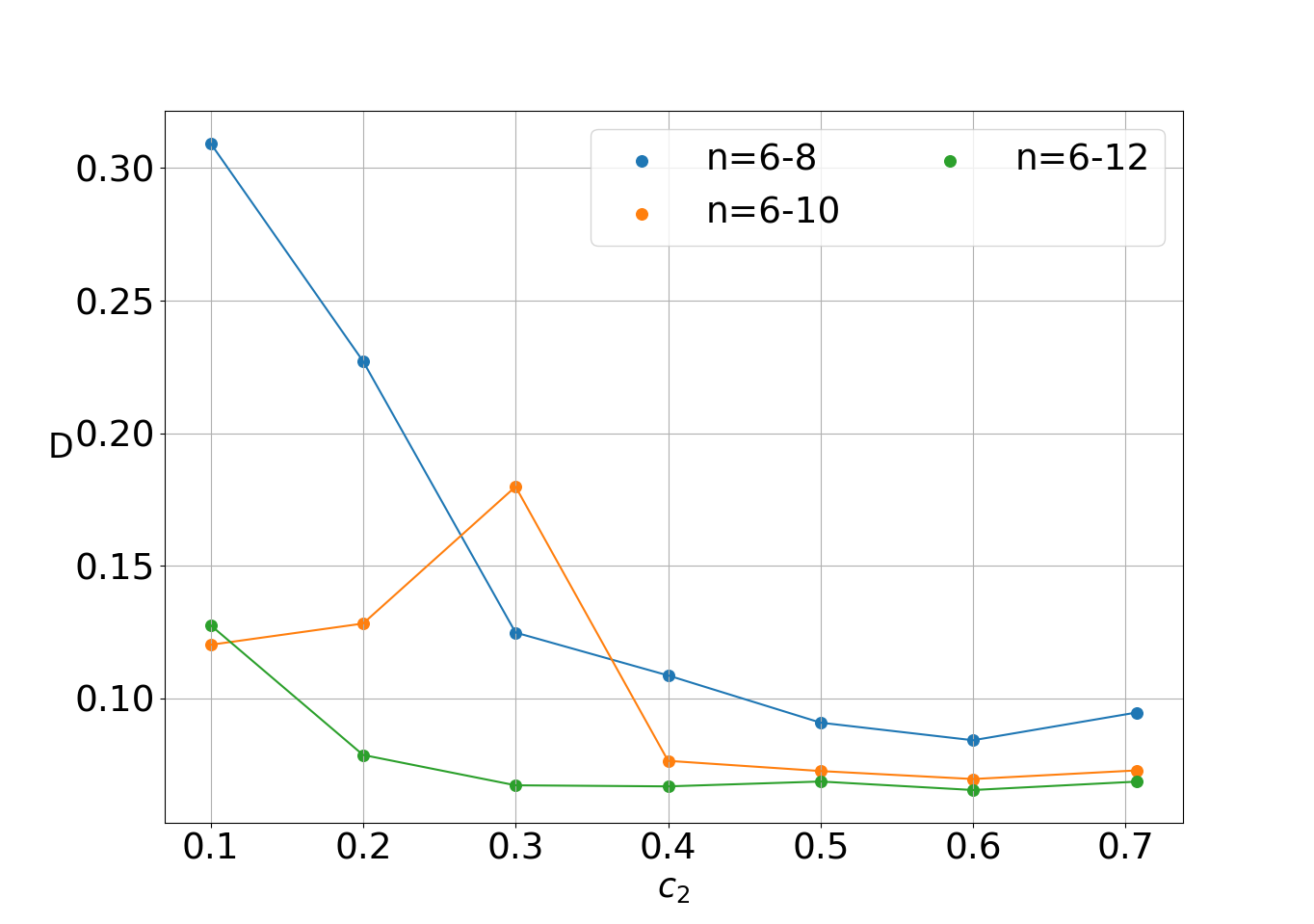}
\caption{Frobenius distance between the distribution of points of a chaotic trajectory with $x_0=0.1, y_0=0.4$ of the truncated coherent states and the corresponding trajectory of the full qubit state for various degrees of entanglement ($c_2$).}\label{sigrisi6}
\end{figure}

This is shown in Fig.~\ref{sigrisi6} where we have a figure similar to Fig.~\ref{sigrisi0}, {but we have ommited the contributions of the low energy levels $0, 1, 2, 3, 4, 5$. In this figure we gradually cover the range of high energies starting at $n_{in}=6$ up to $n_f=8,10,$ and $12$ correspondingly}. The general behaviour is similar to that of Fig.~\ref{sigrisi0}, namely: (a) as $n_f$ increases the value of  the Frobenius distance between  the distribution of points of a chaotic trajectory in the truncated and the non-truncated case $D$ decreases or remains almost constant beyond $c_2=0.3$ The limiting values of $D$ for the same values of the upper $n_f$ when $c_2\to 0.707$ are about the same as in Fig.~\ref{sigrisi0}. (b) For a fixed value of $c_2$ the values of $D$ for increasing $n_f$ (upper) decrease in general and tend to zero.

%

\section{COHERENT STATES WITH COMMON SMALL AMPLITUDES}

A well known fact is that two arbitrary coherent states are not in general orthogonal, but have an interference which decreases with the increase of $a_0$. Up to now we have used the value $a_0=2.5$ and  we had a small interference of order $10^{-5}$. On the other hand, if $a_0$ is small   our model deviates from a two-qubit system, since we can not define the two basis states of the qubits.

\begin{figure}[H]
\centering
\includegraphics[scale=0.35]{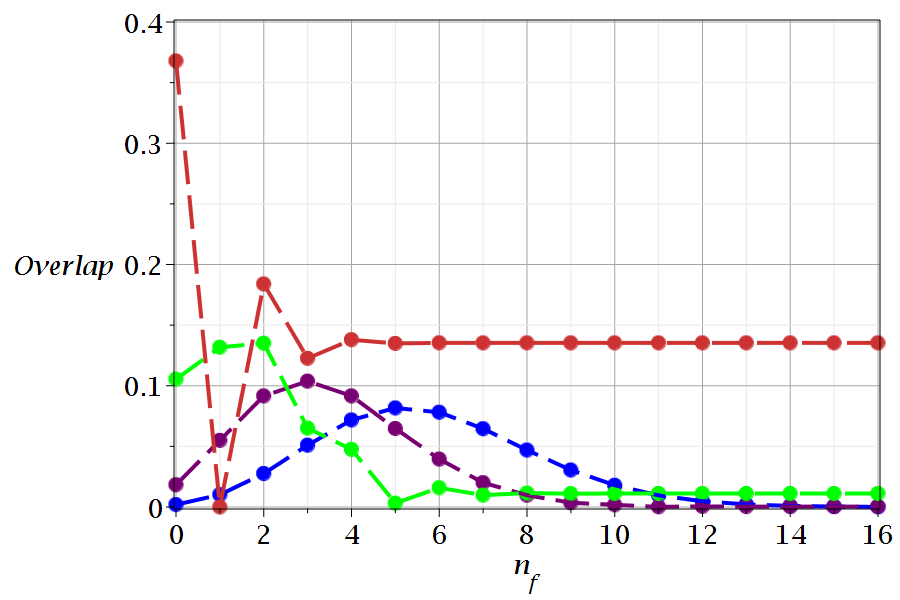}
\caption{The overlap $\int_{-\infty}^{\infty}Y_L(x)Y_R(x)dx$ for various values of the truncation order $n_f$ and for: $a_0=2.5$ (blue), $a_0=2.0$ (purple),$a_0=1.5$ (green), $a_0=1.0$ (red), }\label{orth}
\end{figure}

In fact if $a_0$ is small there is an appreciable overlapping between the two coherent states, as shown in Fig.~\ref{orth} for various values of $a_0$ and of the truncation order $n_f$. The overlap is given by the integral $\int_{-\infty}^{\infty}Y_L(x)Y_R(x)dx$. We observe the abrupt increase of the  interference between $Y_L$ and $Y_R$ for  $a_0<1.5$. In the case  $a_0=1.0$ the qubit model breaks down completely.

A small value of $a_0$ in the non-truncated system implies a small oscillation range for the Gaussian blobs around the origin. In fact the two blobs are formed around the points
\begin{align}\label{Lf}
x_c=\sqrt{\frac{2}{\omega_x}}a_0\cos(\omega_xt), \quad y_c=\sqrt{\frac{2}{\omega_y}}a_0\cos(\omega_yt)
\end{align}
and their maximum distances from the origin are
\begin{align}
d_{max}=\frac{\sqrt{2(\omega_x+\omega_y})a_0}{\sqrt{\omega_x\omega_y}}.
\end{align} 
E.g. for $\omega_x=1, \omega_y=\sqrt{3}$, we find ${d_{max}\simeq 1.78a_0}$.  

\begin{figure}[H]
\centering
\includegraphics[scale=0.17]{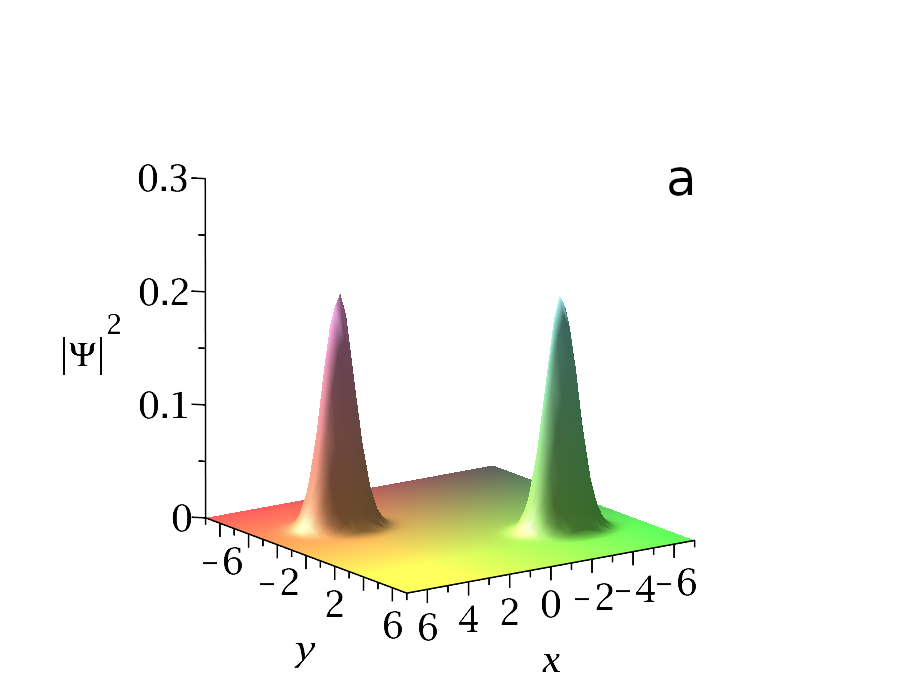}
\includegraphics[scale=0.17]{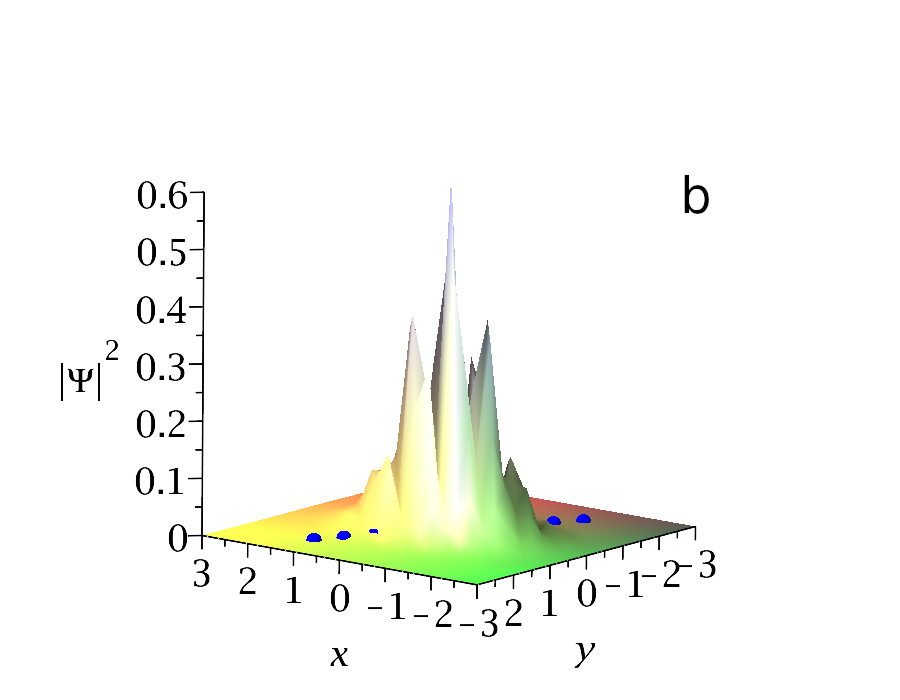}\\
\includegraphics[scale=0.17]{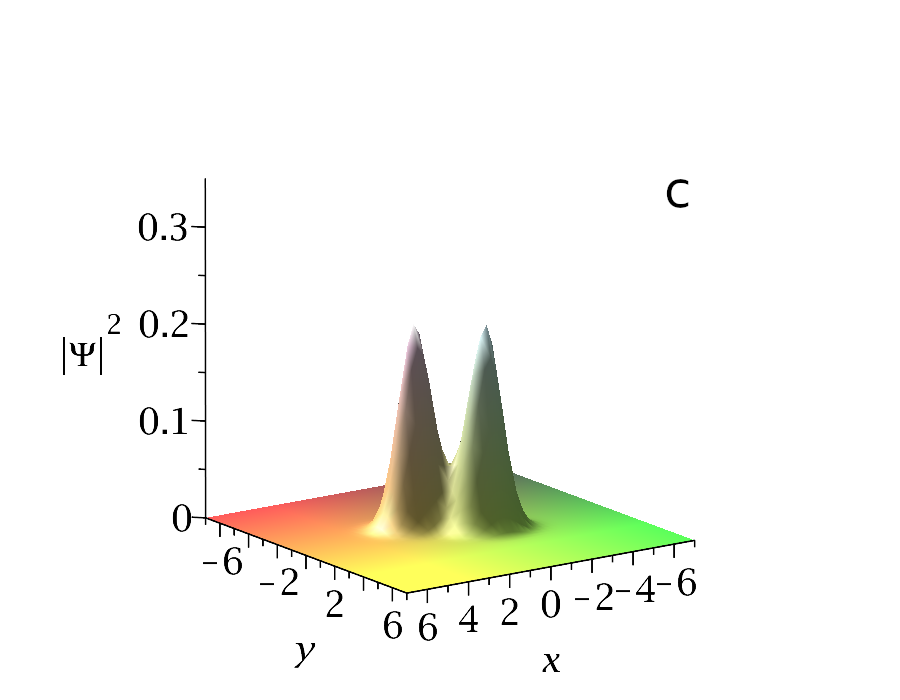}
\includegraphics[scale=0.17]{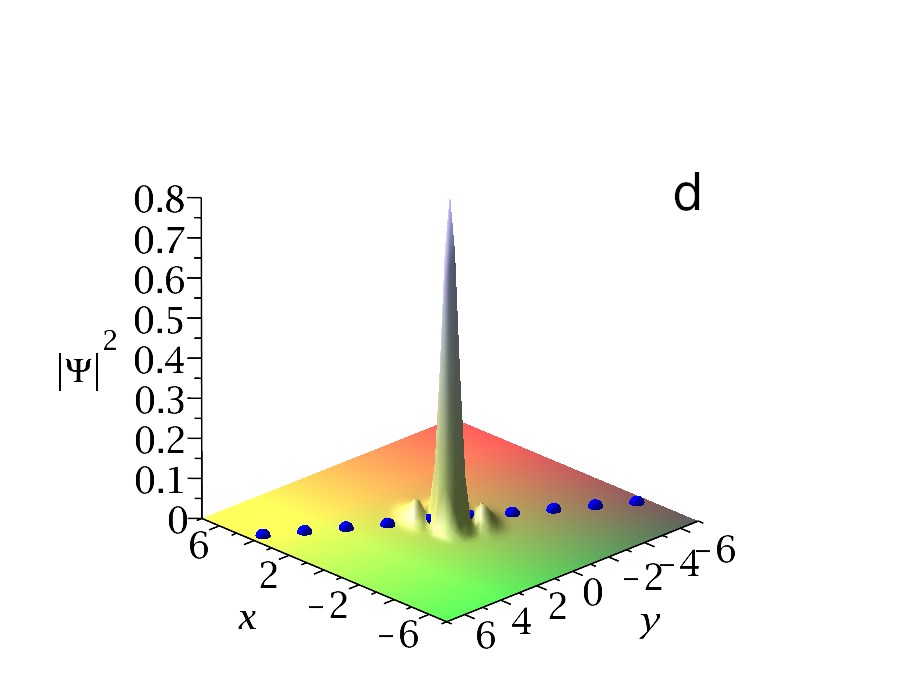}\\
\includegraphics[scale=0.17]{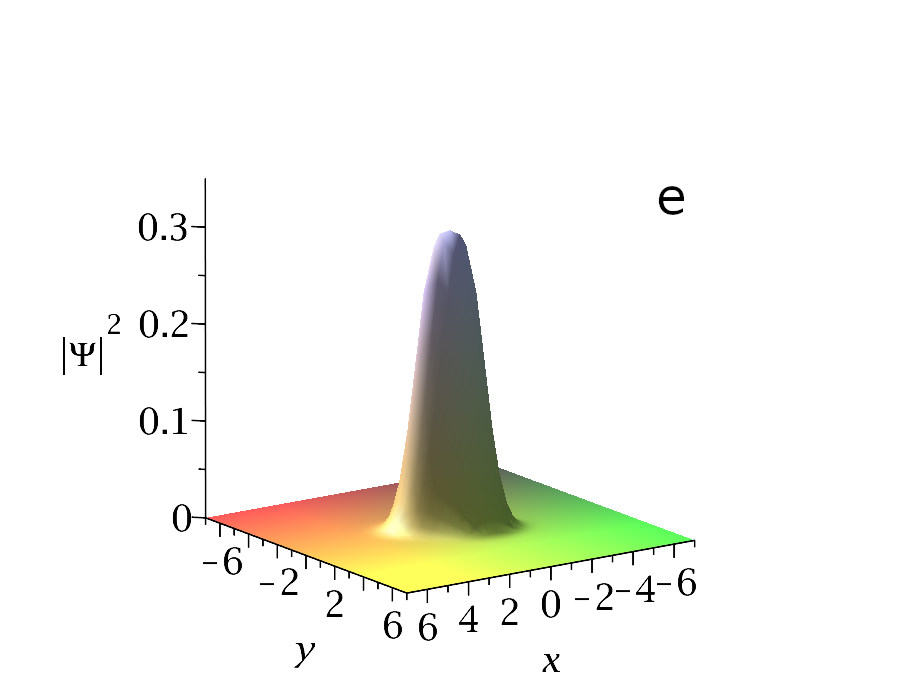}
\includegraphics[scale=0.17]{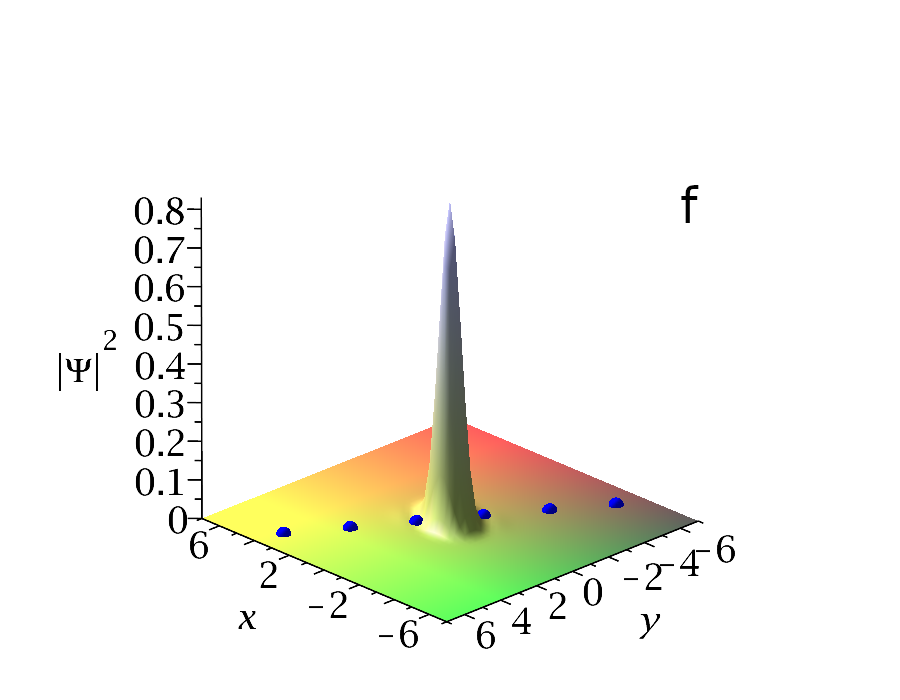}
\caption{$|\Psi|^2$ in the case where $n_f=\infty$ and  $a_0=2.5$ (top figures), $a_0=0.8$ (middle figures) and $a_0=0.5$ (bottom figures) for $t=0$ and $t=4.58$ (collision time). As $a_0$ decreases the nodal points move away from the central blob (the center of collision). }\label{nq1}
\end{figure}

In Figs.~\ref{nq1} we plot the probability density $|\Psi|^2$ for $t=0$ and $t=4.58$ (collision time)  of the maximally entangled ($c_2=\sqrt{2}/2$) state when  $a_0=2.5$ (top panel), $a_0=0.8$ (middle panel) and  $a_0=0.5$ (bottom panel). The mean energy values in these cases are $a_0^2=6.25$, $a_0^2=0.64$ and $a_0^2=0.25$ correspondingly. As $a_0$ decreases the two blobs approach the origin (Figs.~\ref{nq1}a,c) and for $a_0=0.5$ the two blobs have almost joined near the origin (Fig.\ref{nq1}c). In the last case the two blobs overlap strongly  and they can never be  separated. 

\begin{figure}[H]
\centering
\includegraphics[scale=0.25]{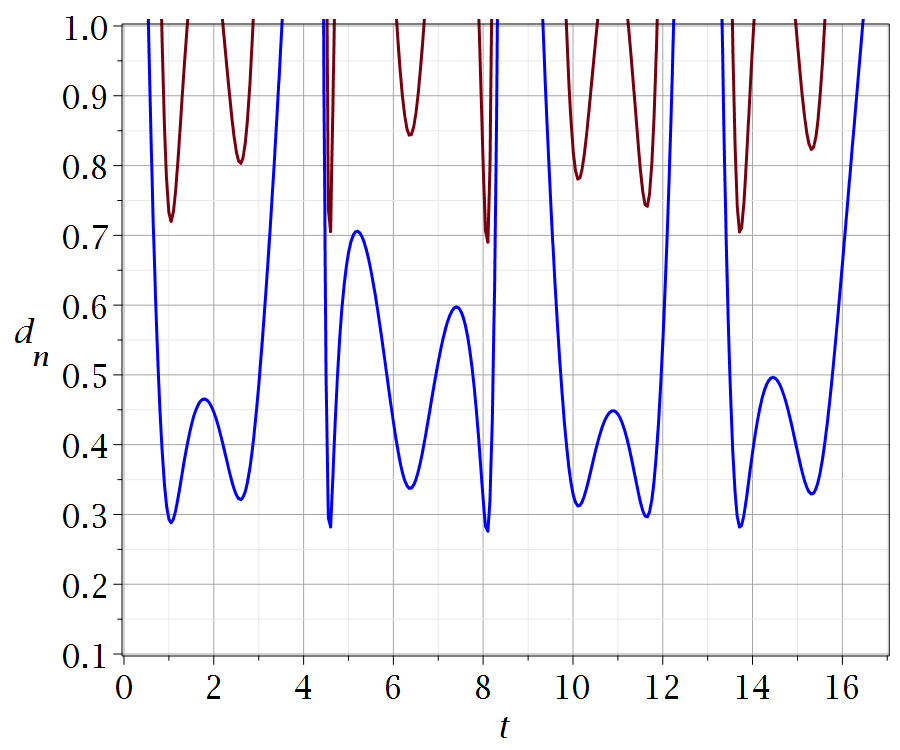}
\caption{The distance between the origin and the nodal points $k=\pm 1$ as functions of time for $c_1=c_2=\sqrt{2}/2$, when  $a_0=2.5$ (blue curve) and $a_0=1$ (burgundy curve). The greater the amplitude $a_0$ the smaller the distance $d_n$.}\label{npd}
\end{figure}

On the other hand the nodal points ($\Psi_{Real}=\Psi_{Imag}=0$) are at the positions 

\begin{eqnarray}
\nonumber\label{xnod}&x_{n}={\frac {\sqrt {2}
\left( k\pi\,\cos \left( 
\omega_{y}\,t \right) +\sin \left( 
\omega_{y}\,t \right) \ln  \left( 
\left| {\frac {c_{1}}{c_{2}}} \right|  
\right)  \right) }{4\sqrt {\omega_{x}}a_{0}\,\sin \left(  
\omega_{xy}  t \right) } },\\&
\label{ynod}y_{n}={\frac {\sqrt {
2} \left(k\pi\, \cos \left( \omega_{x}t 
\right) +\sin \left( \omega_{x}t \right) 
\ln  \left(  \left| 
{\frac {c_{1}}{c_{2}}} \right|  \right)  
\right) }{4\sqrt {\omega_{y}}a_{0}\,\sin 
\left( \omega_{xy}\,t \right) }},
\end{eqnarray}
with $k\in Z $, $k$ even for $c_1\cdot c_2<0$
or odd for $c_1\cdot c_2>0$, and $\omega_{xy}\equiv \omega_x-\omega_y$. For $c_1=c_2$ their {minimum distances} from the center are:
\begin{align}
d_n=\frac{\sqrt{2}\pi}{4a_0\sqrt{\omega_x\omega_y}\sin(\omega_{xy}t)}\sqrt{\omega_x\cos^2(\omega_xt)+\omega_y\cos^2(\omega_yt)}
\end{align}
When $t=0$ the nodal points are at infinity, but during a collapse they approach the origin (Figs.~\ref{nq1}b,d,f).
The minimum values of $d_n$ for $|k|=1$ are given in Fig.\ref{npd} and they are inversely proportional to $a_0$.

\begin{figure}[H]
\centering
\includegraphics[scale=0.27]{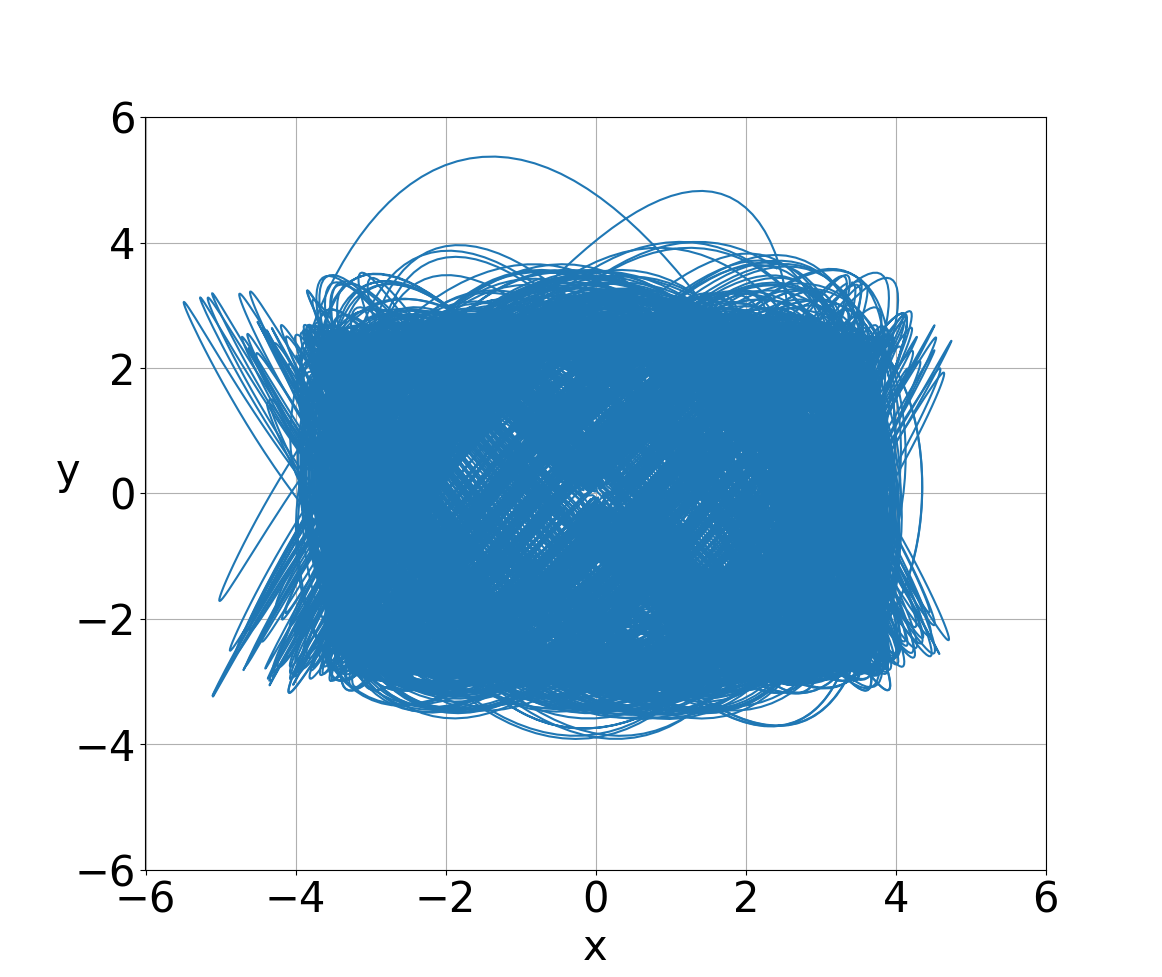}
\caption{A typical chaotic trajectory in the case $a_0=2.5$ up to $t=4000$ ($c_2=\sqrt{2}/2, x_0=3, y_0=0$).}\label{chaot25}
\end{figure}

\begin{figure}[H]
\centering
\includegraphics[scale=0.2]{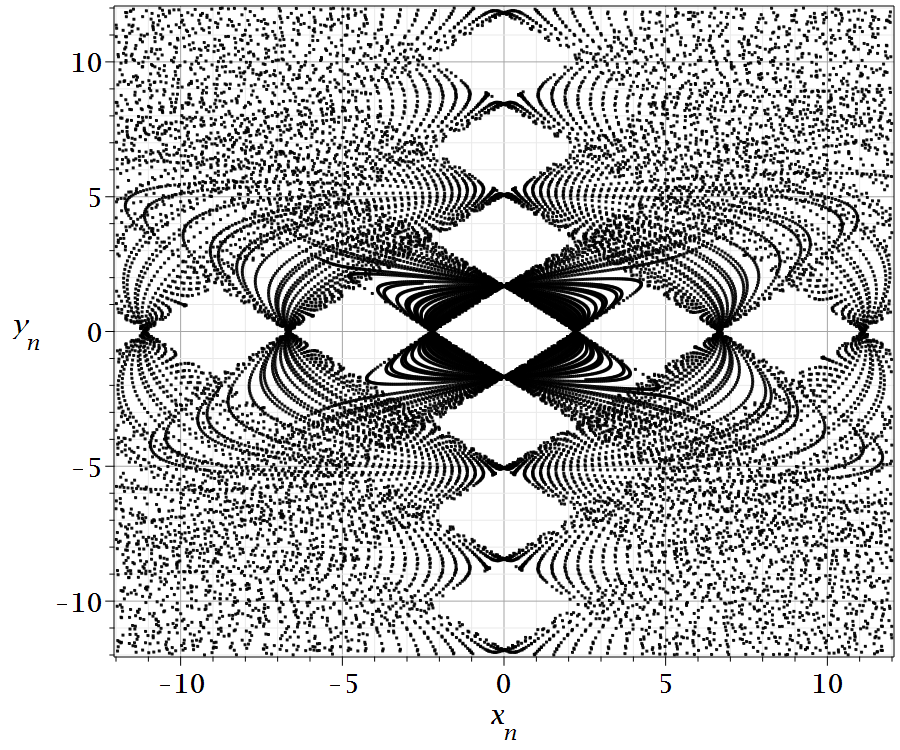}a
\hspace{0.7cm}
\includegraphics[scale=0.2
]{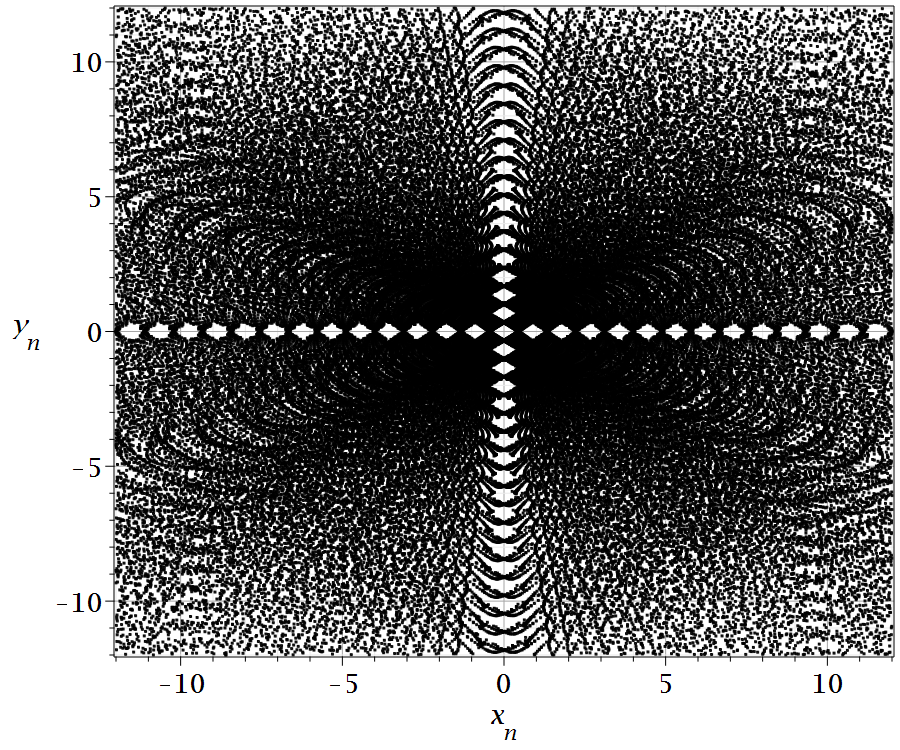}b
\caption{Nodal trajectories of the non-truncated system in the cases  $a_0=0.5$ (a) and $a_0=2.5$ (b). The nodal trajectories have several empty regions which decrease as $a_0$ increases. For $a_0\to\infty$ the empty regions practically disappear. }\label{nt}
\end{figure}

\begin{figure}[H]
\centering
\includegraphics[scale=0.25]{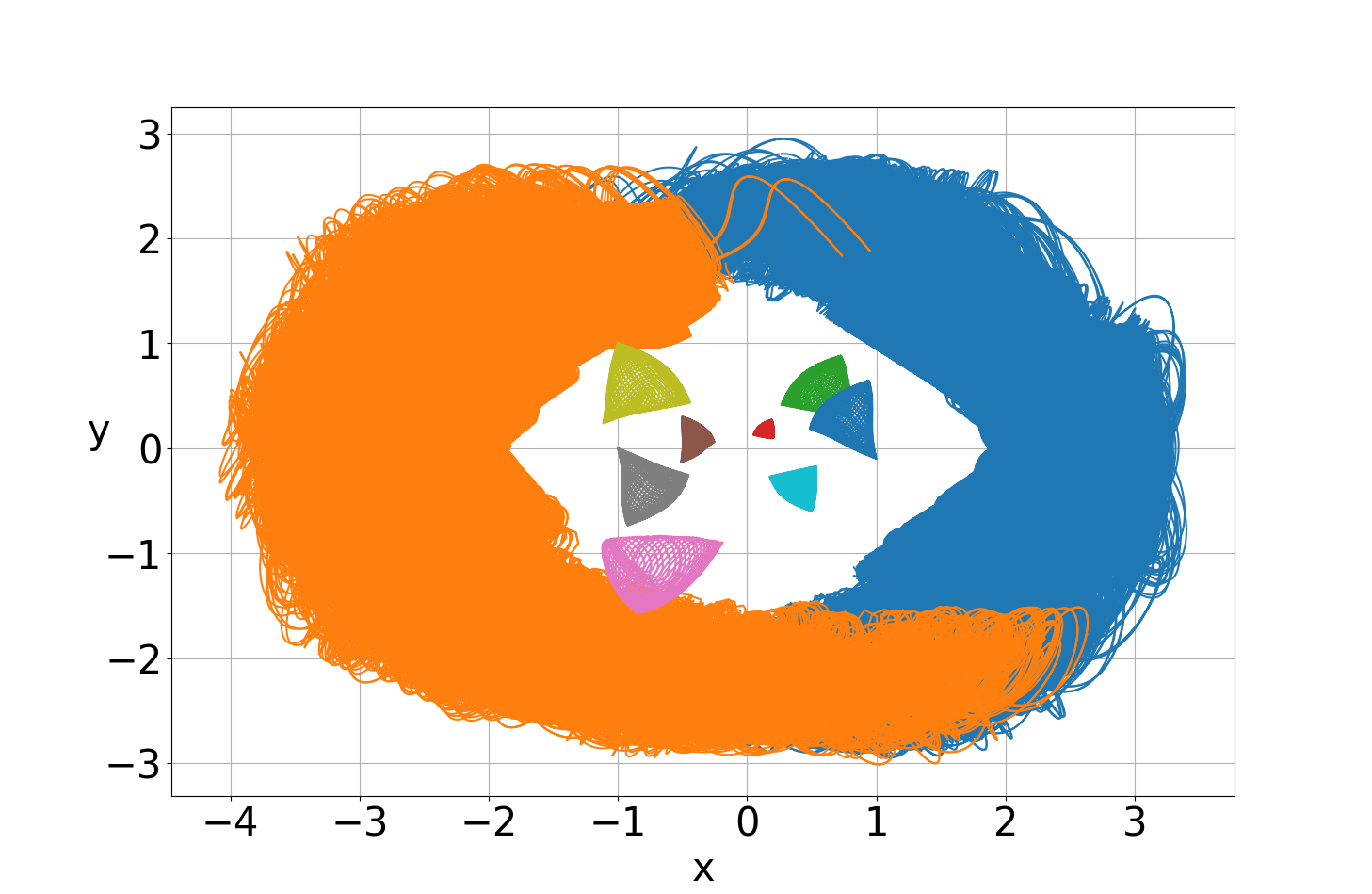}
\caption{Bohmian trajectories of the full coherent state system in the case $a_0=0.5$. The chaotic trajectories are integrated up to $t=4\times 10^5$. We observe that the central region of the configuration space around the origin is now dominated by ordered trajectories (integrated up to $t=200$). This is due to the fact that the nodal points do not enter in the central region.}\label{plir_org}
\end{figure}

In the case $a_0=2.5$ the minimum values of $d_n$ are about $d_n\simeq 0.3$, i.e. they are much smaller than the initial distances of the blobs from the center ($d_{max}=10$). In general, for large $a_0$ the nodal points  come closer to the origin than the tops of the two blobs during a collision, therefore they scatter the trajectories close to the origin, which become chaotic. An example of such a chaotic trajectory is given in Fig.~\ref{chaot25}. However, for small $a_0$ the nodal points are outside the blobs and they do not produce chaotic trajectories near the center. In fact, the trajectories of the nodal points in the coherent state for $a_0=0.5$ leave empty a large area around the center (Fig.~\ref{nt}a) and further areas above, below and on the right and  on the left. These empty regions are much larger in the case $a_0=0.5$ (Fig.~\ref{nt}a) than in the case $a_0=2.5$ (Fig.~\ref{nt}b). As a consequence, for $a_0=0.5$ the trajectories that start close to the origin ($d<1$) do not approach any nodal point, therefore they are ordered. In Fig.~\ref{plir_org} we see the areas covered by several ordered trajectories around the origin.

\begin{figure}[H]
\centering
\includegraphics[scale=0.32]{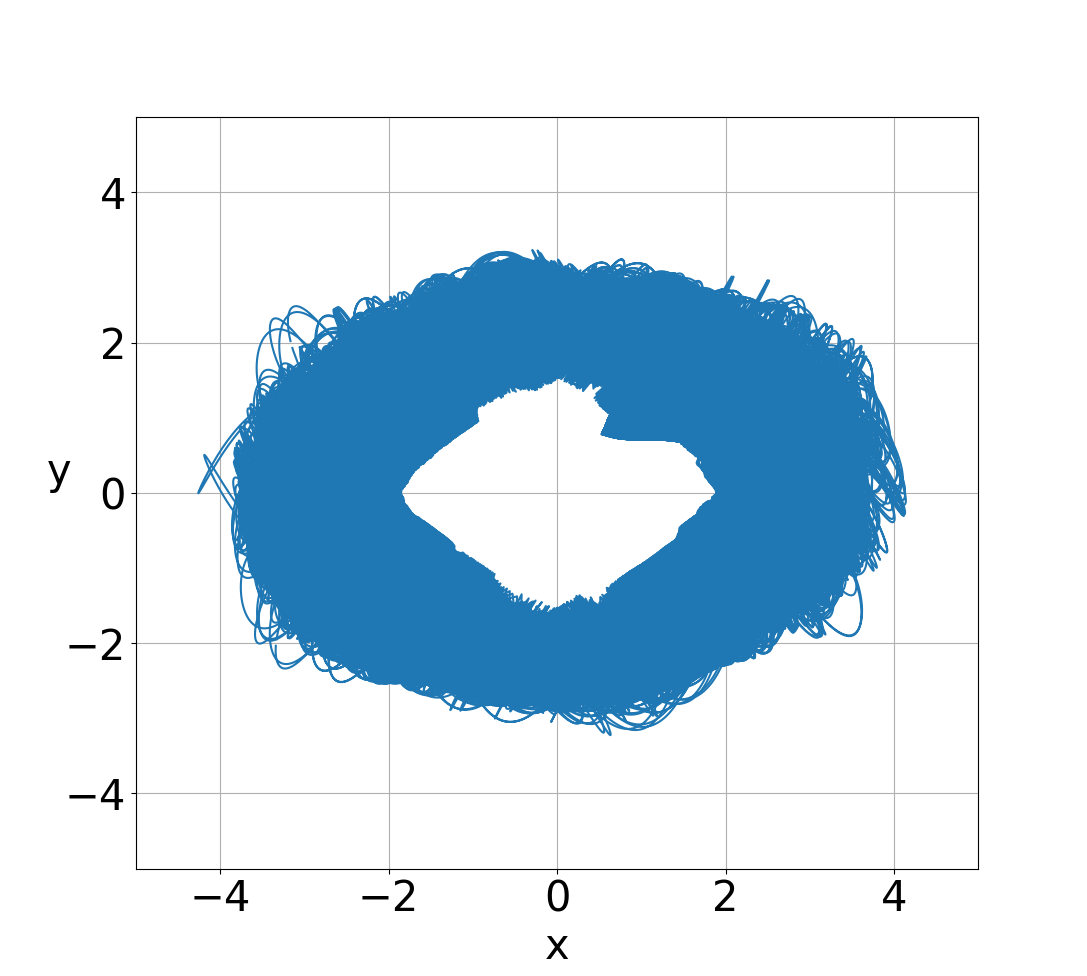}
\caption{A chaotic trajectory ($x_0=3, y_0=0$) covering a ring around the central blob of $|\Psi|^2$ in the case $a_0=0.5$, up to $t=7\times 10^5$ .}\label{donut}
\end{figure}

\begin{figure}[H]
\centering
\hspace{-2cm}\includegraphics[scale=0.17]{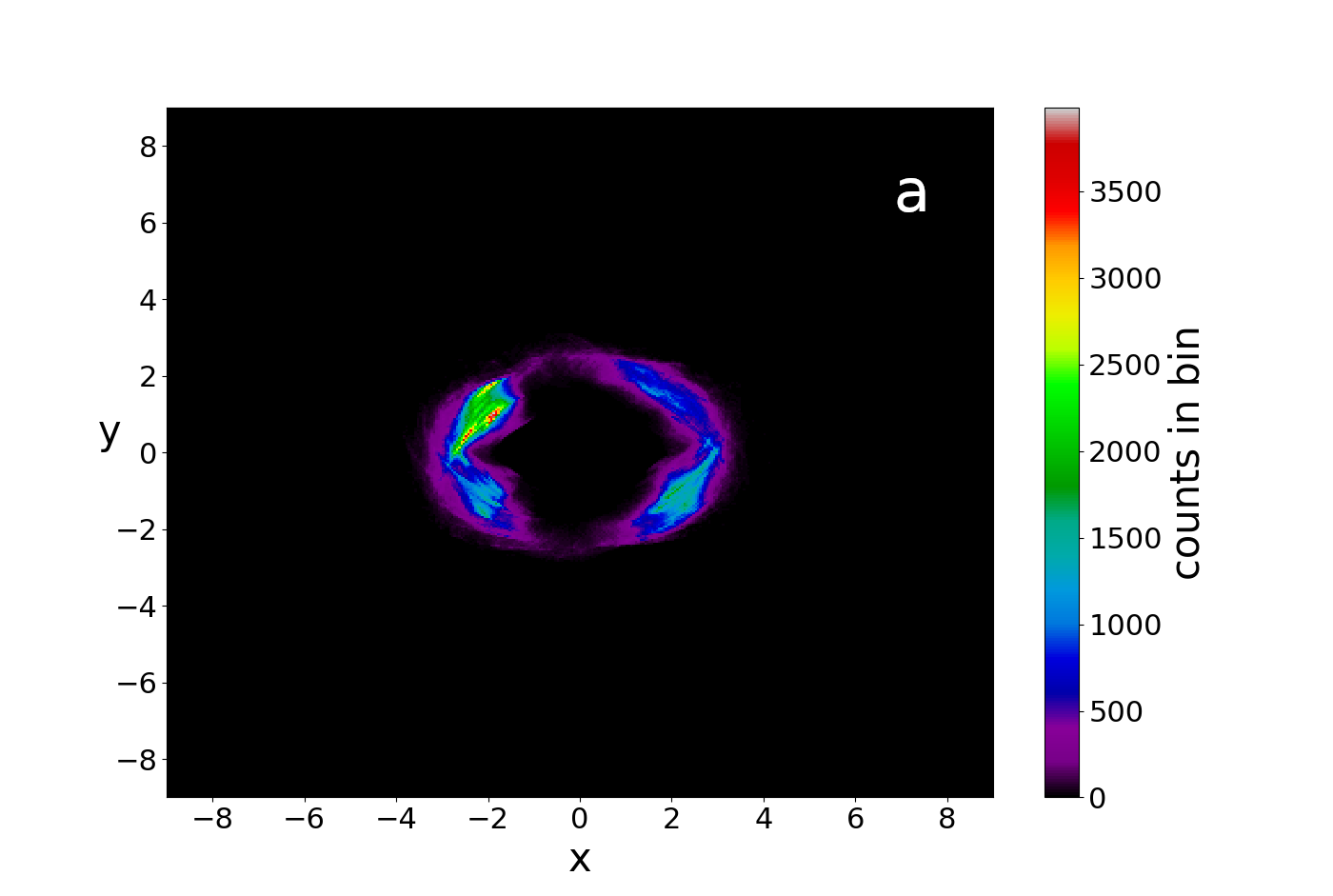}
\includegraphics[scale=0.17
]{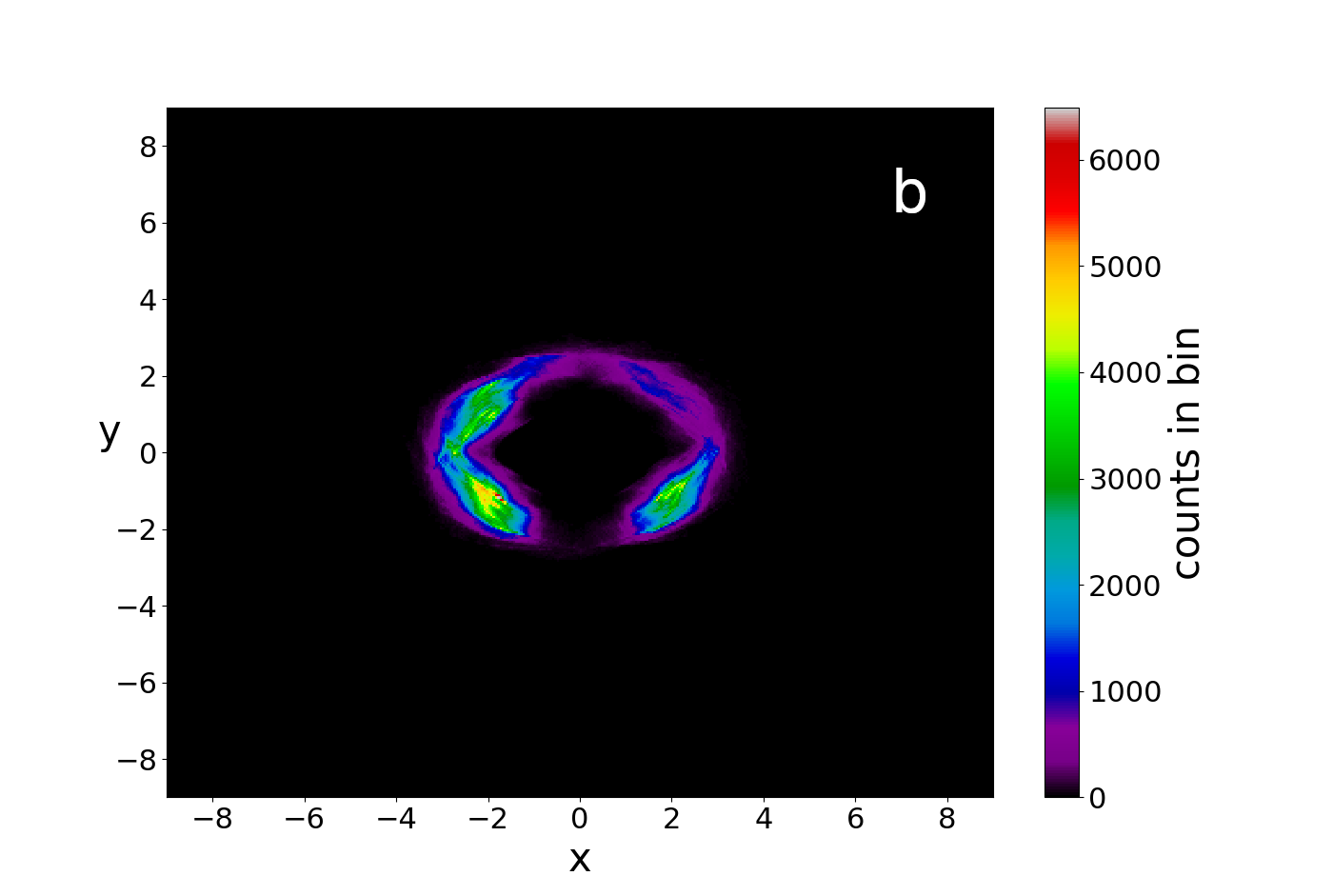}\\\hspace{-2cm}
\includegraphics[scale=0.17
]{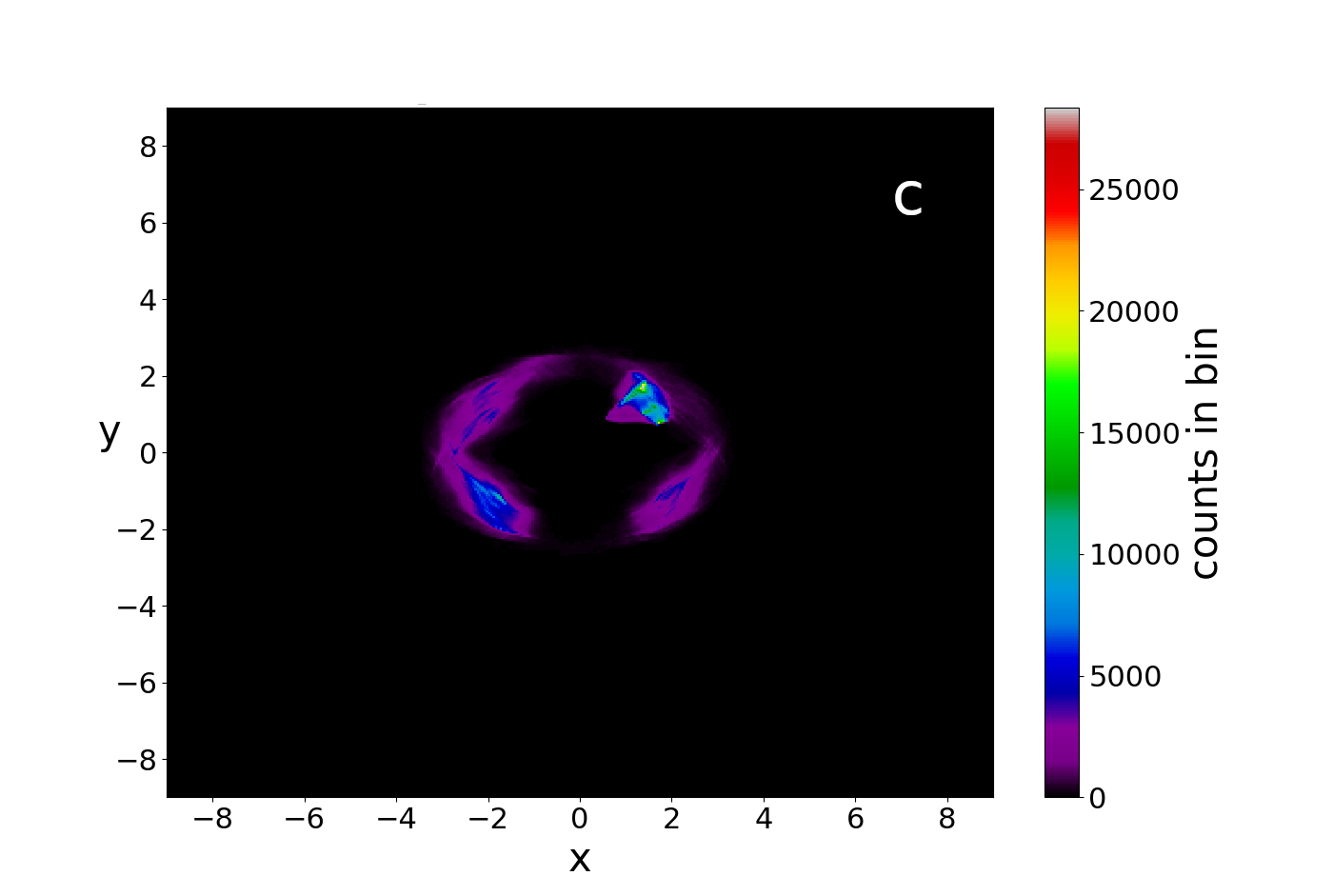}
\includegraphics[scale=0.17
]{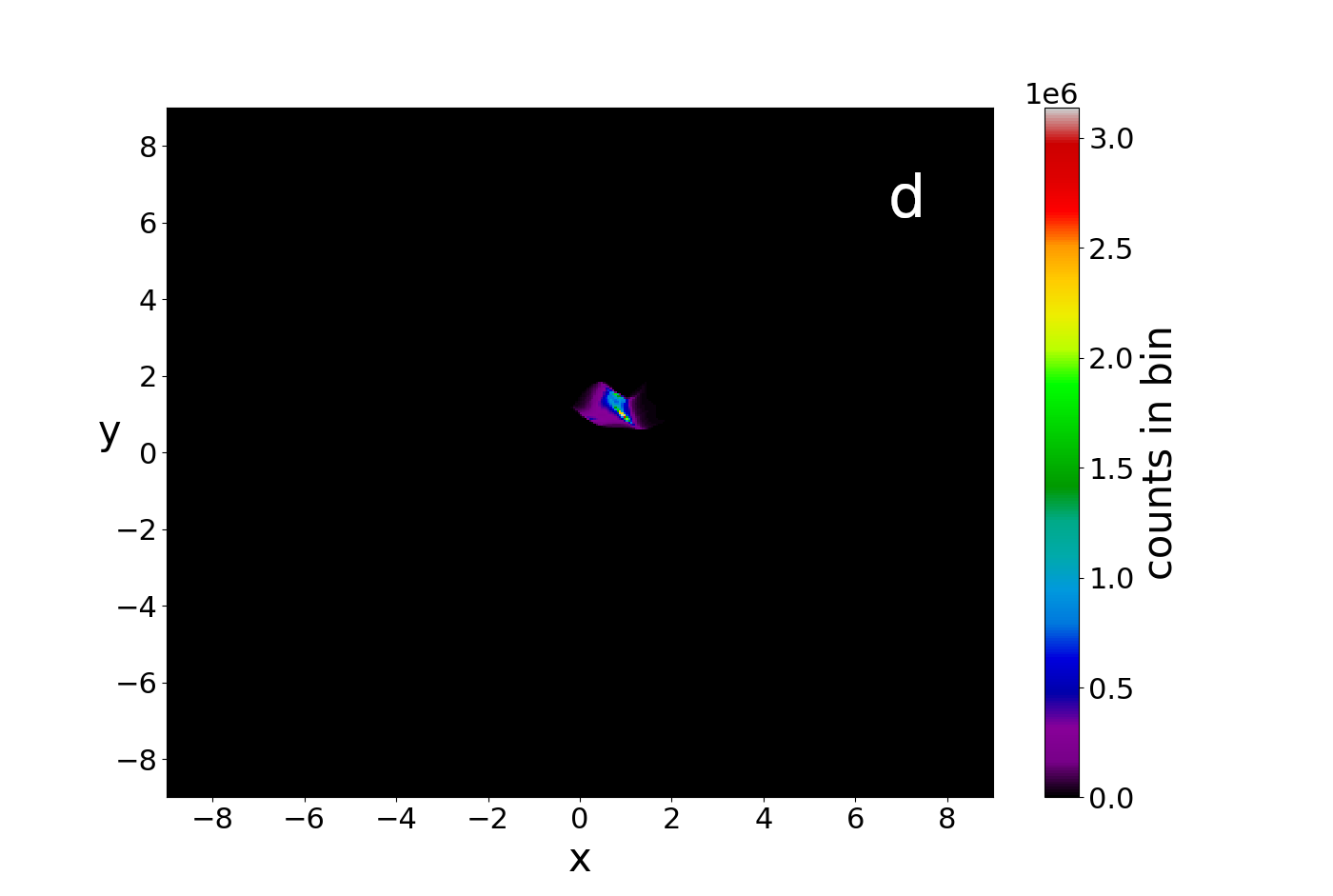}
\caption{Successive colorplots of the points of the distribuition of a chaotic trajectory $x_0=3, y_0=0$ in the case $a_0=0.5, c_2=\sqrt{2}/2$ for a) $t=200000$, b) $t=400000$, c) $t=600000$ and d) $t=10^7$.}\label{colf}
\end{figure}

On the other hand, the  trajectories starting further away from the origin are deflected by the NPXPCs and become chaotic. In Fig.~\ref{plir_org} we see two chaotic trajectories. Each of them covers, after a time $t=4\times 10^5$, about half of a thick ring surrounding the origin.  After a time $t=7\times 10^5$ both trajectories of Fig.~\ref{plir_org} cover the whole ring (Fig.~\ref{donut}), but not in  a uniform way. In fact, the points of the trajectories stay most of the time in  rather small parts of the configuration space, although from time to time they cover the whole ring (Fig.~\ref{colf}a,b,c,d). That was verified for times up to $t=10^7$. 

\begin{figure}[H]
\centering
\includegraphics[scale=0.25]{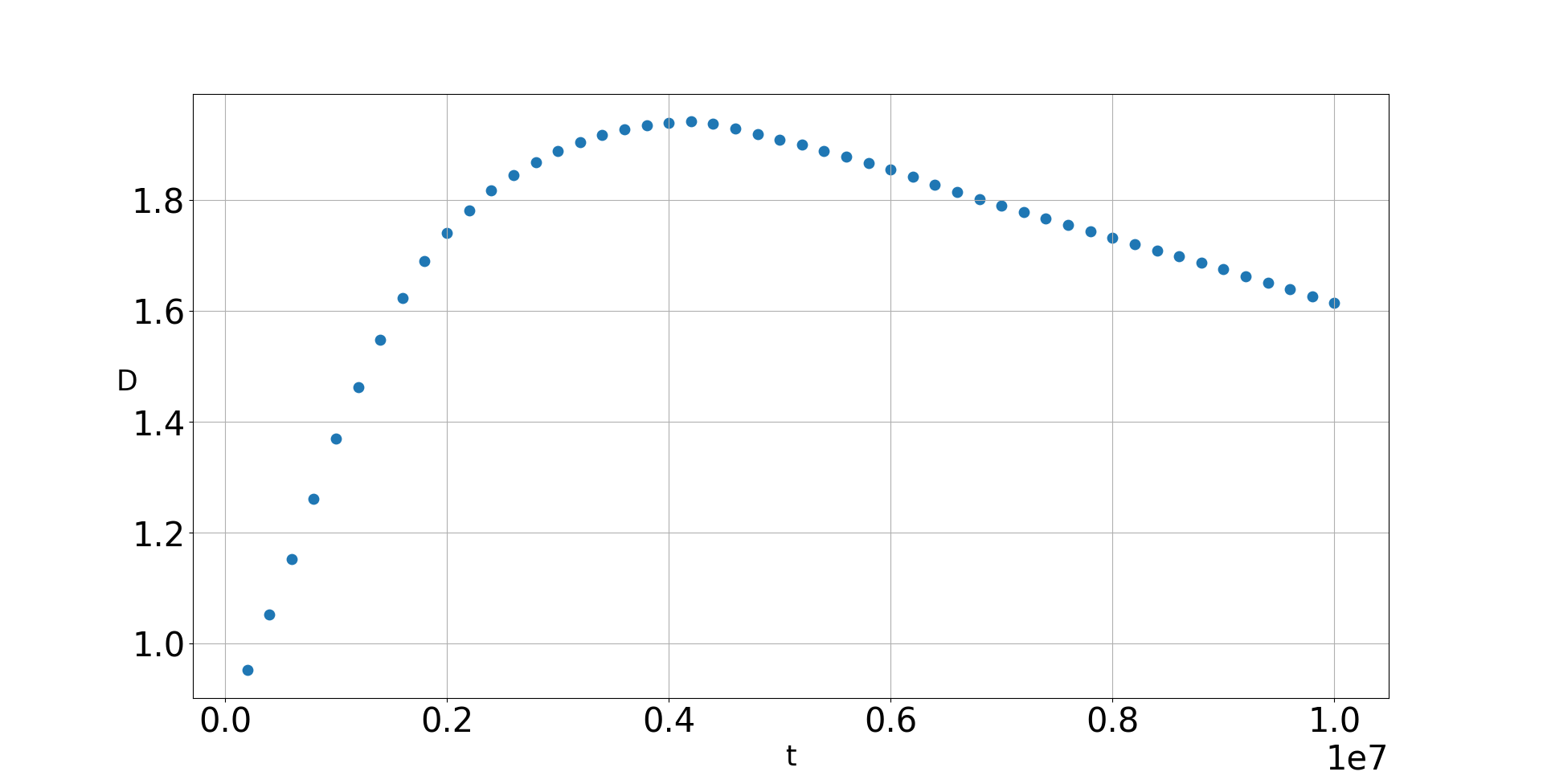}
\caption{The Frobenius distance between two chaotic trajectories forming rings around the main blob, in the case $a_0=0.5$. $D$ increases  up to $t=4\times 10^6$. Only then it starts to decrease linearly in time with a power law, according to which $D$ will reach zero at $t\simeq 4\times 10^7$.}\label{fd}
\end{figure}

 Therefore these trajectories do not appear to be  ergodic  for long times. This is also shown by calculating the Frobenius distance $D$ between the distributions of the  points of two such chaotic trajectories as a function of time (Fig.~\ref{fd}). However, a rough extrapolation of Fig.~\ref{fd} shows that $D$ approaches zero beyond $t=4\times 10^7$. Thus  the chaotic trajectories become ergodic after an extremely long time.

The problem now is how the trajectories for large $a_0$ (Fig.~\ref{chaot25}) transform to those of Fig.~\ref{plir_org} with the decrease of the amplitude $a_0$.

%

\begin{figure}[H]
\centering
\hspace{-2cm}
\includegraphics[scale=0.18]{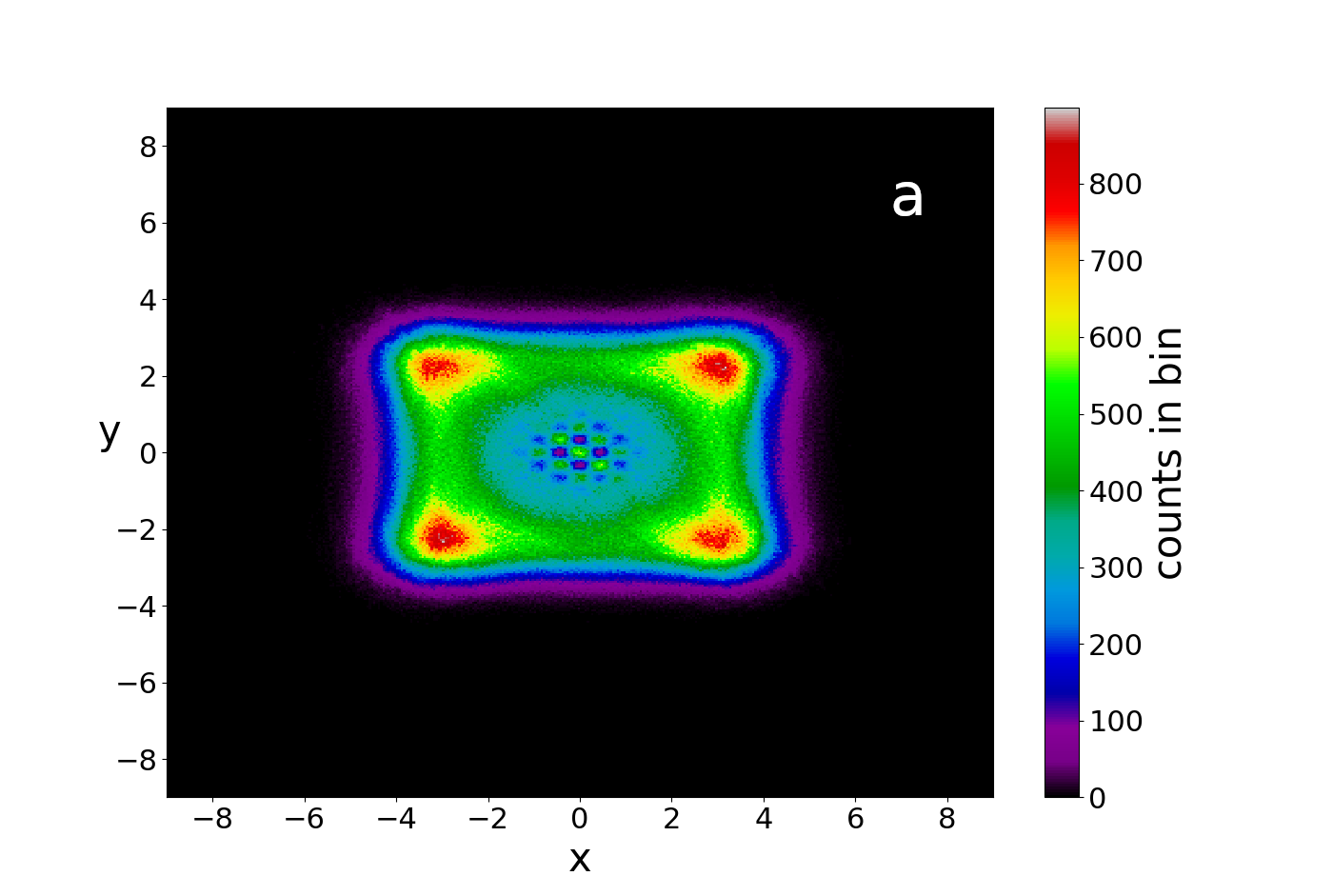}
\includegraphics[scale=0.18]{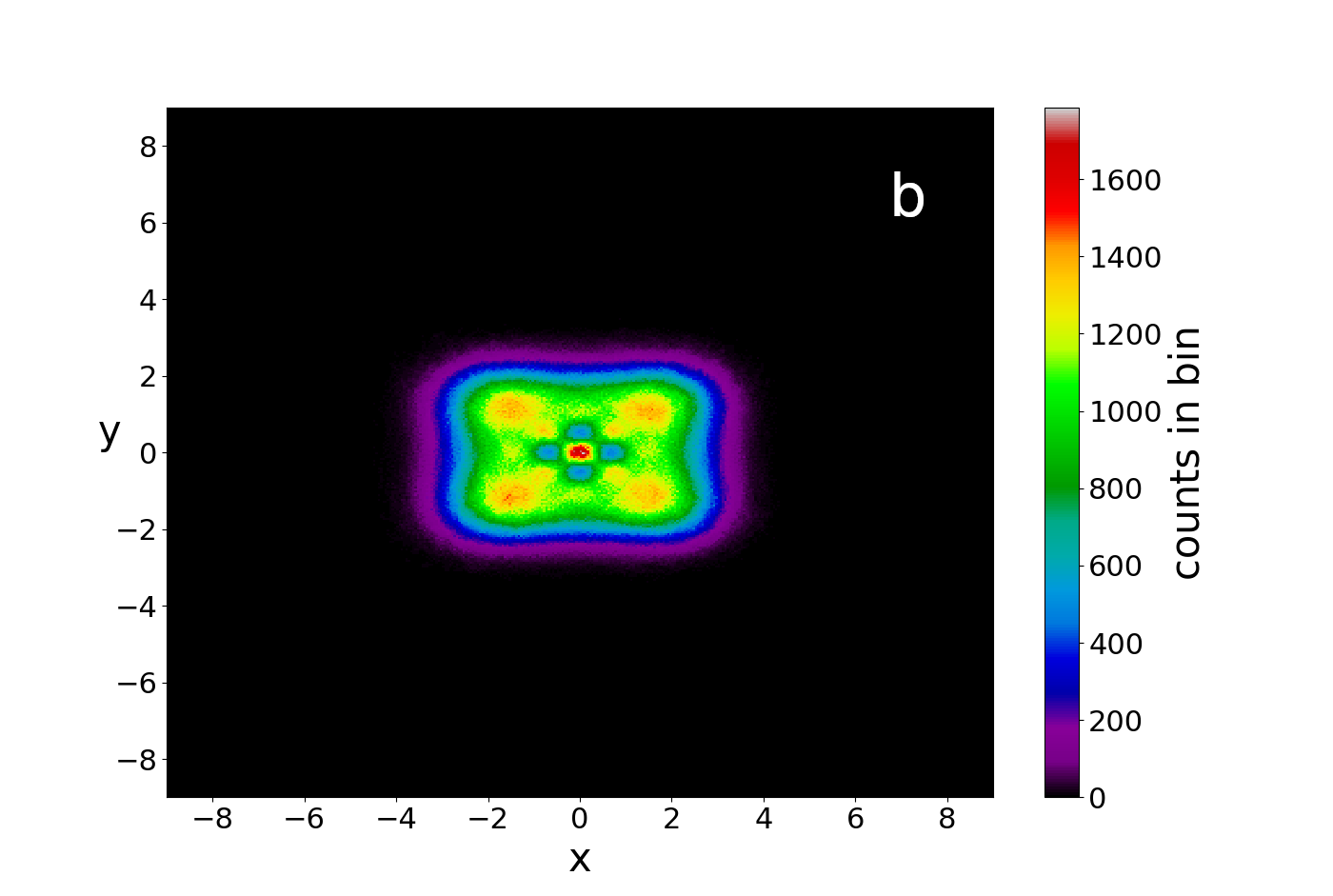}\\\hspace{-2cm}
\includegraphics[scale=0.18]{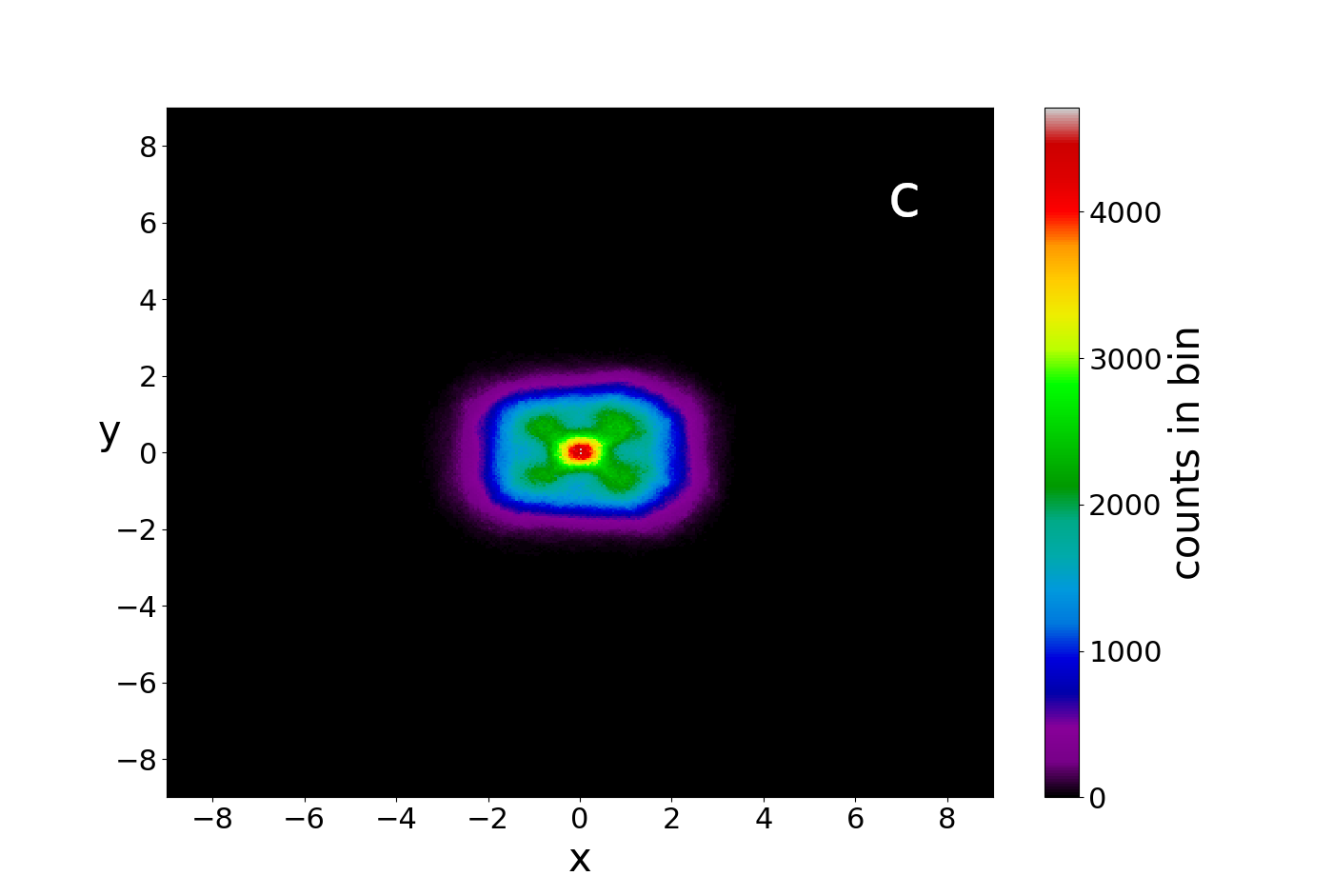}
\includegraphics[scale=0.18]{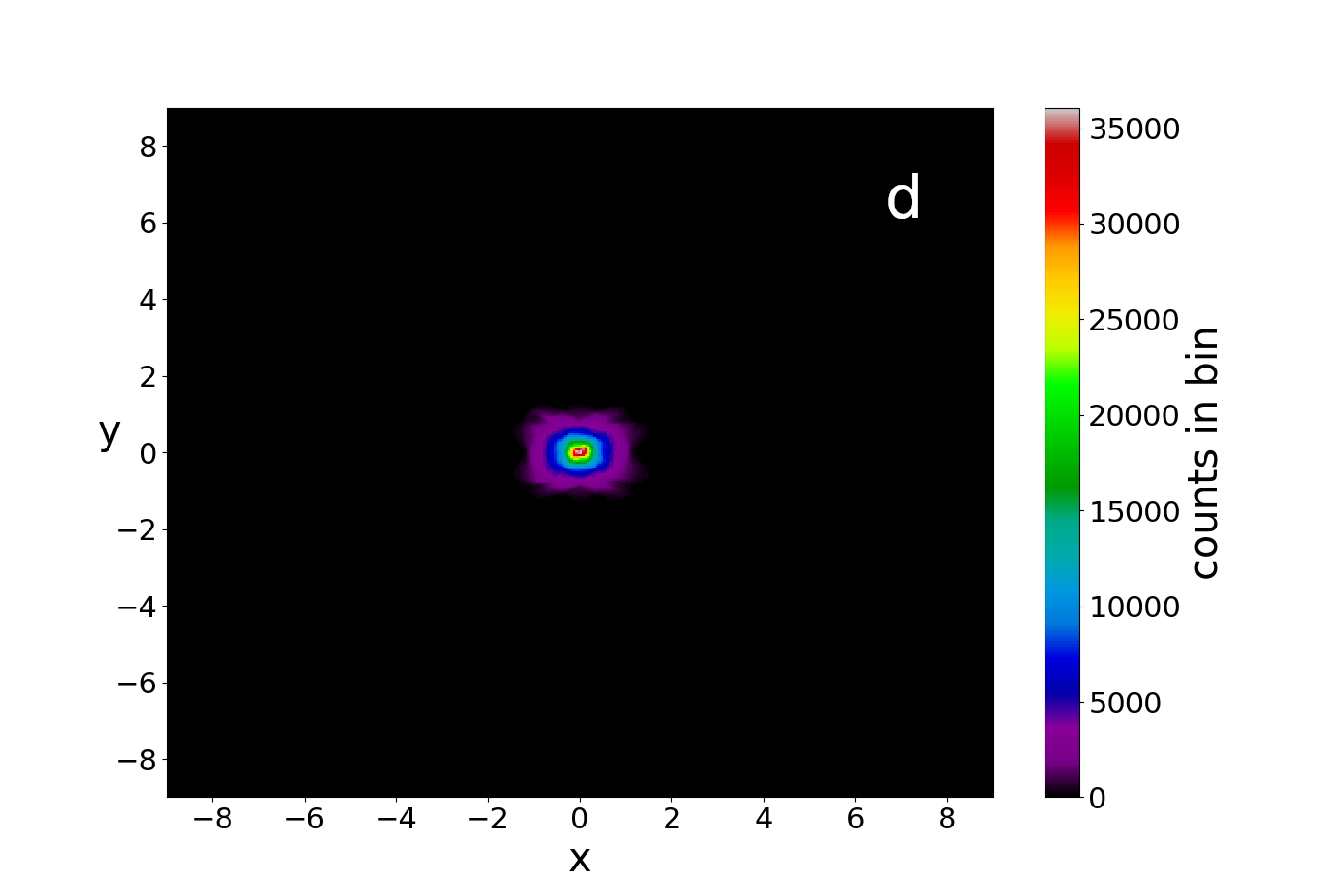}
\caption{Colorplots of trajectories with the same initial conditions ($x_0=0.1, y_0=0.2$) in the maximally entangled state and for $t=10^5$ , when:  (a) $a_0=2.5$ , (b) $a_0=1.5$, (c) $a_0=1$  and (d) $a_0=0.8$. We observe how the 4 discrete blobs tend to join as $a_0$ decreases.}\label{cps}
\end{figure}

In Fig.~\ref{cps} we construct the colorplots giving the distributions of the points of the chaotic trajectories, that start near the center of the configuration space. We observe that, for large $a_0$, these plots form the usual pattern of four bright blobs near the corners of the support of the wavefunction (Fig.~\ref{cps}a). As $a_0$ decreases the distances of the bright blobs and the whole space distribution become smaller (Fig.~\ref{cps}b, c) and at the same time a maximum of density (red) appears at the center. For even smaller $a_0$ (Fig.~\ref{cps}d), the four blobs have joined the central maximum, which now dominates the distribution. However, the  trajectories starting further away from the center form rings, as in the case of Fig.~\ref{donut}. As $a_0$ decreases the region near the center that contains regular trajectories becomes larger (Fig.~\ref{nt}).

\begin{figure}
\centering
\hspace{-2cm}
\includegraphics[scale=0.18]{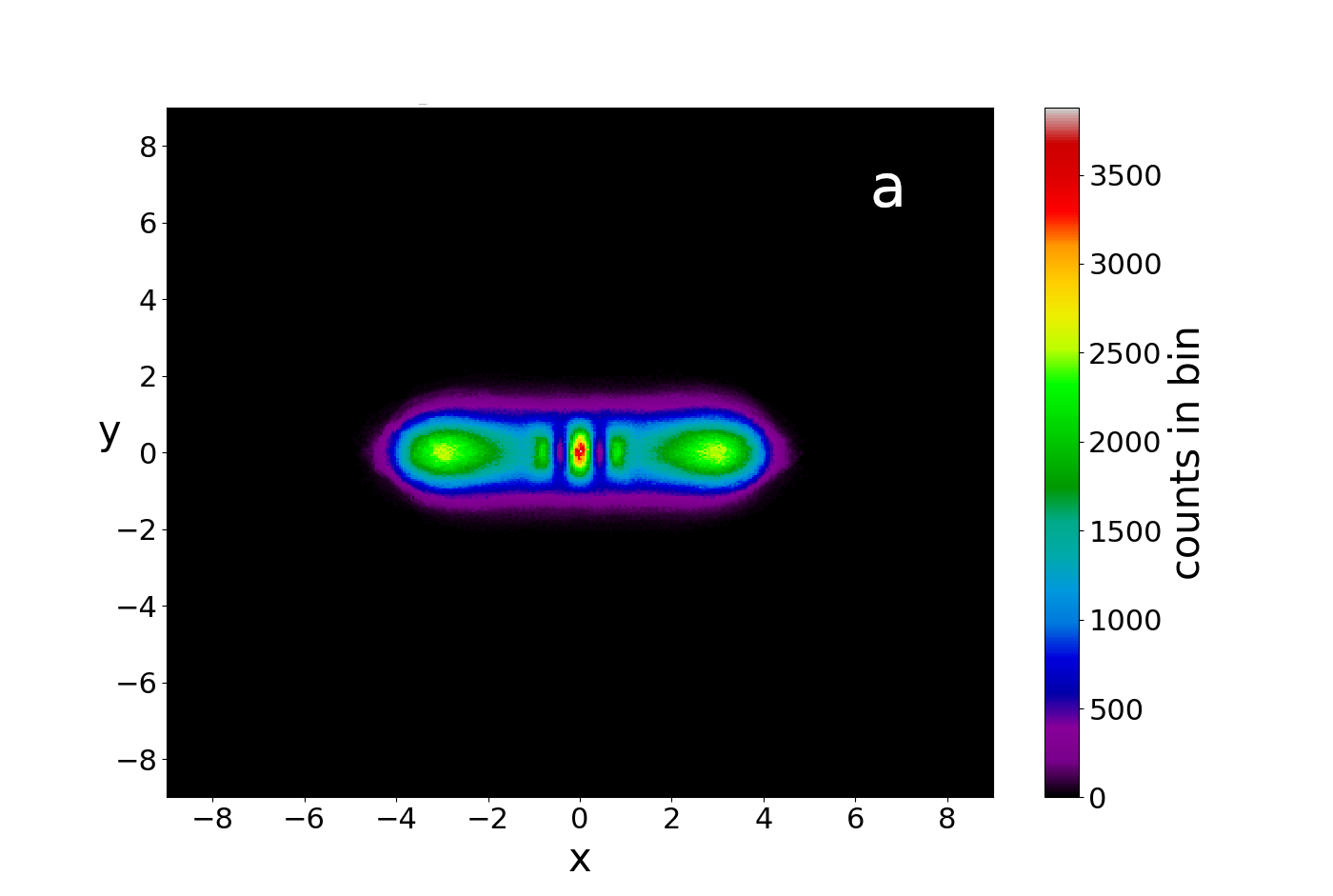}
\includegraphics[scale=0.18]{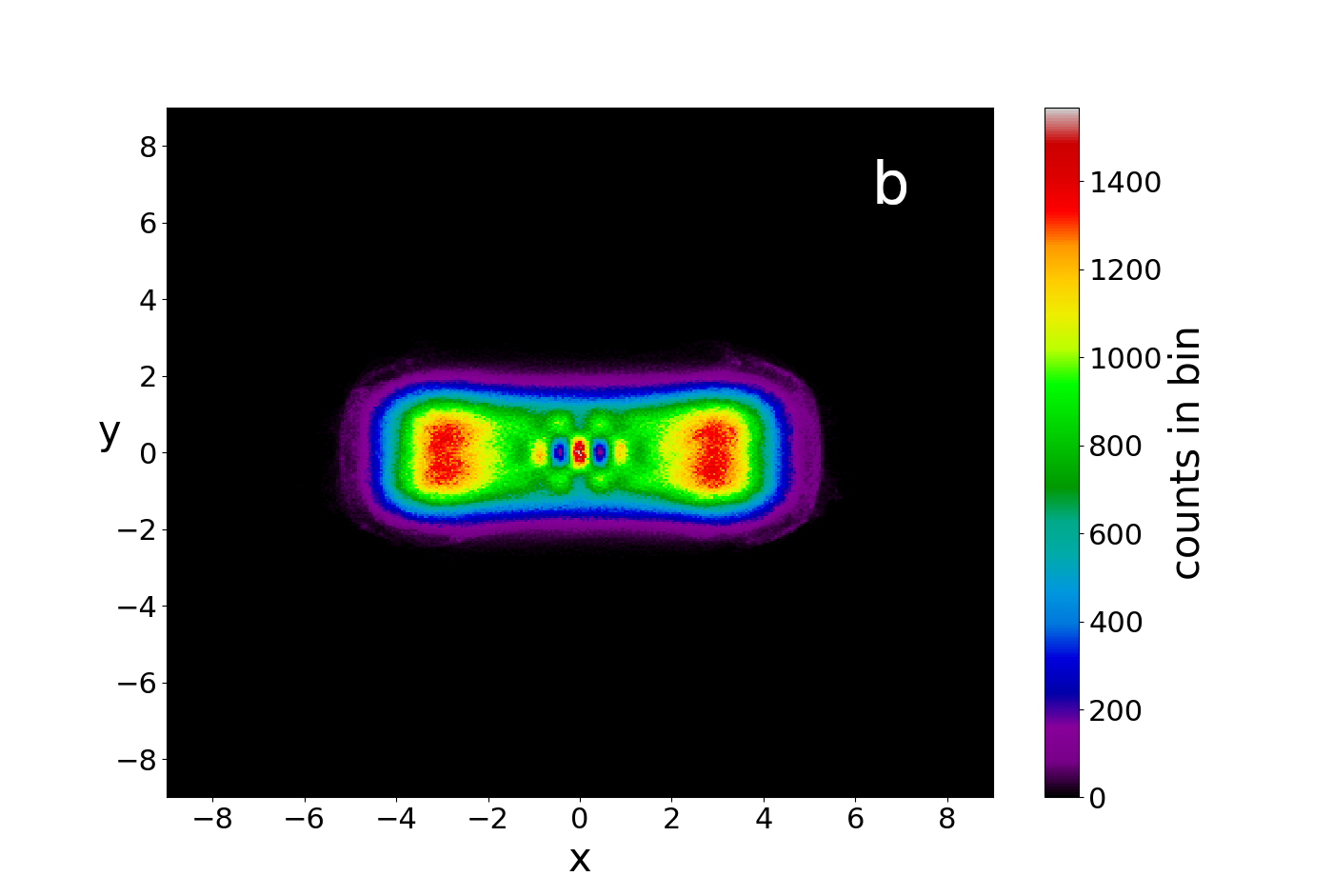}
\caption{Colorplots of trajectories with the same initial conditions  ($x_0=0.1, y_0=0.2$) in the maximum entangled state and up to $t=10^5$ for $a_0=2.5$ and (a) $b_0=0.5$, (b) $b_0=1$.}\label{a0b0}
\end{figure}

Up to now we have considered coherent states with the same amplitudes $a_0$ along $x$ and $y$ coordinates. In Figs.~\ref{a0b0}a,b we consider the limiting distribution of the points of the trajectories in cases where the entanglement is maximum and the amplitude along $x$  is $a_0=2.5$ but the amplitude along $y$ is smaller, e.g. $b_0=0.5$ (Fig.~\ref{a0b0}a) and $b_0=1$ (Fig.~\ref{a0b0}b). By comparison with the the non-truncated qubit case $a_0=b_0=2.5$ (Fig.~\ref{cps}a), we observe that, as $b_0$ decreases, the pattern becomes more concentrated close to the $x$-axis. In particular, in Fig.~\ref{a0b0}b the two upper maxima of the probability density come closer to the lower maxima (in comparison with Fig.~\ref{cps}a), and in Fig.~\ref{a0b0}a the upper and the lower maxima have practically joined, so that we have only two maxima, one on the left and one on the right. On the other hand, the distances between the maxima along the $x$-axis remain about the same. We have checked numerically that all the chaotic trajectories are ergodic and their number increases with entanglement, as in the case with common amplitudes.


%


\section{TRUNCATED COHERENT STATES WITH SMALL AMPLITUDES}
In the cases where the amplitudes are small, we saw 
that the probability density is dominated by a central 
blob whose motion is confined very close to the origin. 
\begin{figure}[H]
\centering
\includegraphics[scale=0.35]{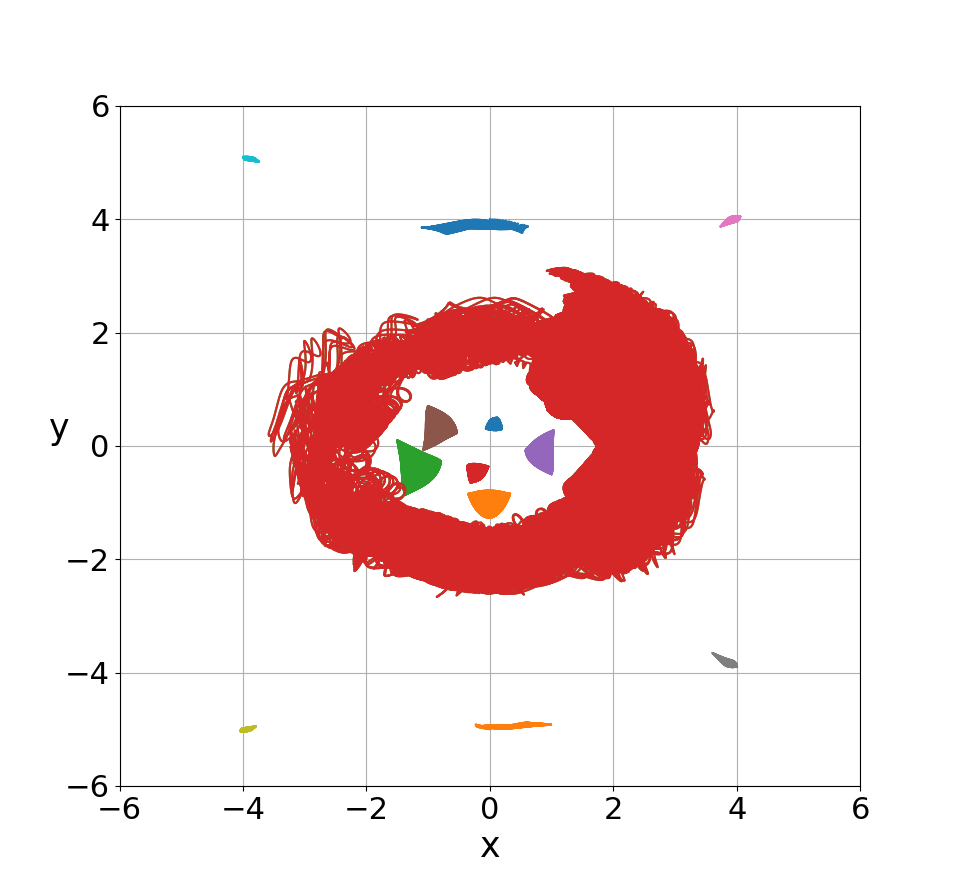}
\caption{Several ordered trajectories in the truncated case with $a_0=b_0=0.5$ inside and outside the support (approximately $x\in[-1.85, 1.85]$ and $y\in[-1.5, 1.5]$), when the truncation is $n_f=2$ for times up to $t=200$. We  also show a typical chaotic trajectory surrounding the support of the wavefunction for times up to $t=4\times 10^5$.}\label{mat}
\end{figure}
Thus we expect that even with a truncated wavefunction 
we will find mostly ordered trajectories in that 
region, as shown in Fig.~\ref{mat}. These trajectories are of different forms, but they all have zero Lyapunov characteristic number \cite{Contopoulos200210}. There are also many chaotic trajectories around the central blob of the probability density $|\Psi|^2$ which  form rings around the central blob, as the red trajectory of Fig.~\ref{mat}. These trajectories surround the support of $\Psi$, whose extent is approximately equal to the central hole of Fig.~\ref{mat}. However, due to the truncation of the energy levels, i.e. the finite number of nodes, we find also ordered trajectories further  outside the support of the wavefunction. This is the basic difference between the truncated and the non truncated case, where all the ordered trajectories are inside the support of $\Psi$ \cite{tzemos2021role}.

We must note that, for a detailed study of ordered and chaotic  Bohmian trajectories, one needs to compute the trajectories of the 
nodal points. However, it is not possible, in general, 
to solve analytically the algebraic equations defining 
the positions of the nodal points in the truncated cases, 
which are of degree $2n_f$. Thus we are obliged to find these points numerically. This calculation is difficult  when the various nodal points approach each other and when their velocities become large (when they go to or come from infinity).

\begin{figure}[H]
\centering
\includegraphics[scale=0.37]{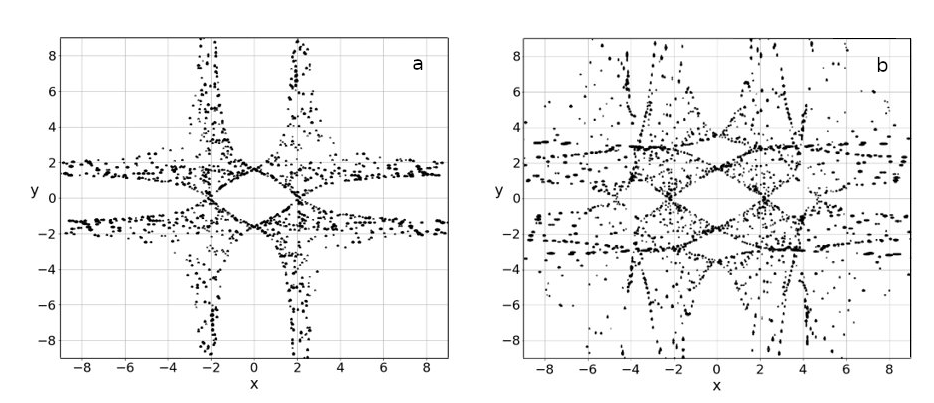}
\caption{The dots represent the very small contour plots around the nodal points in the case where $a_0=b_0=0.5$ , for  $t\in[0,100]$ with time step $\Delta t=0.01$ and $n_f=2$ (a), and for $t\in[0,50]$ and  $n_f=4$, with time step $\Delta t=0.025$ (b).}\label{komboi_ap}
\end{figure}

Nevertheless, we managed to make a 
graphical depiction of the nodal points of the truncated coherent 
states by use of contour plots. Namely, in Fig.~\ref{komboi_ap} 
we exploit the fact that the  velocity of a Bohmian particle becomes extremely large 
as we approach the nodal points, and plot the contour plots of 
the Bohmian flow when the velocity is very high ($|v|=500$). 
These contours are very small closed curves around the nodal 
points. This graphical method is, of 
course general (i.e. works for every wavefunction) and is 
useful since  it gives approximately the positions of the 
nodes in the configuration space for any time. 
In Figs.~\ref{komboi_ap}a,b we show the contour plots in 
cases with $a_0=b_0=0.5$ and $n_f=2$  up to time $t=100$ with time step 
$\Delta t=0.01$ (Fig.~\ref{komboi_ap}a) and  $n_f=4$ up 
to time $t=50$ with time step $\Delta t=0.025$ (Fig.~\ref{komboi_ap}b).\footnote{As the truncation energy corresponding to $n_f$ increases 
the time of the calculation of these figures increases 
significantly due to the high degree of the polynomials 
involved in the wavefunction.} 


In the case $n_f=2$ we 
have four nodal points whose distribution leaves extended 
empty regions in the configuration space. In the case 
$n_f=4$ we have eight nodal points and their distribution 
covers many empty regions of the case $n_f=2$. As $n_f$ 
increases the empty regions tend to take the form of Fig.~\ref{nt}a.

 A limiting case of a system in which the values of $n$ are truncated from above and below so that only one value remains along $x$ ($n_x$) and $y$ ($n_y$) is the wavefunction $Y_{n_x,n_y}(x,y,t)=Y_{n_x}(x,t)Y_{n_y}(y,t)$. However if we have a sum of such terms $\Psi(x,y,t)=\sum_{n_x,n_y}c_{n_x,n_y}Y_{n_x,n_y}(x,y,t)$ we have in general both chaos and order. Such cases were considered in our previous papers \cite{efthymiopoulos2007nodal,contopoulos2008ordered} and they are similar to the cases considered in the present section. In particular, Figs.~3 of \cite{efthymiopoulos2007nodal,contopoulos2008ordered} are similar to  the present Fig.~\ref{komboi_ap}a.


\section{CONCLUSIONS}

We study the interplay between order and chaos in representative cases of bipartite Bohmian systems, and its implications on the dynamical establishment of Born's rule by an arbitrary initial distribution of Bohmian particles.

 In previous papers we considered the case of two qubits with a wavefunction of the generic form $\Psi=c_1Y_R(x,t)Y_L(y,t)+c_2Y_L(x,t)Y_R(y,t)$ ($|c_1|^2+|c_2|^2=1$), which covers all the range of entanglements, from zero up to maximum entanglement. These qubits are represented by coherent states of two non-interacting quantum harmonic oscillators. The infinite number of energy levels inside the Poisson distribution of the coherent states implied the existence of infinitely many NPXPCs, which scatter the incoming trajectories and produce chaos.  The probability density $|\Psi|^2$ of this model is characterized by two Gaussian blobs which move in the configuration space and collide from time to time \cite{ tzemos2020chaos,tzemos2020ergodicity}. The sizes of these blobs depend on the degree of the entanglement.

The chaotic trajectories of this model were found to be essentially ergodic for any non-zero entanglement, while their number increases with the entanglement. In fact, we found that in the maximally entangled state ($c_2=c_1=\sqrt{2}/2$) all the trajectories are chaotic and ergodic. Then any initial distribution of such trajectories will finally approach Born's rule ($P=|\Psi|^2$) after a long time. However, when the entanglement is small there are also  ordered trajectories close to the center of the leading blob. Thus Born's  rule is not accessible if the initial distribution does not contain approximately the same proportion of ordered and chaotic trajectories with that of the Born distribution. 

In the present paper we study various cases that deviate from the two-qubit model by changing in many ways the physical parameters of the basis states of the qubits. The two major modifications we consider are:

\begin{enumerate}
\item  The truncation  of the coherent states which compose the qubits, at various energy levels for large amplitudes. From a Bohmian perspective this means that we worked with a finite number of NPXPCs.  By plotting the long limit distributions of various truncations and by numerically calculating the motion of the nodal points in two truncated cases we observed the contribution of the various energies in the long limit dynamics of the ideal two-qubit system.  When the number of nodal points is small, the patterns produced by the distribution of the points of the trajectories are different from the pattern of the full state, but as the number of the nodal points increases the patterns approach the pattern of the non-truncated system.

In the truncated cases we find that besides the two main blobs of $|\Psi|^2$ there are also several secondary blobs for all times (and not only during collisions). Moreover we find ordered trajectories, not only close to the center of the leading blob, but also outside the support of the wavefunction, even in the maximum entanglement case.

On the other hand the chaotic trajectories  are still approximately ergodic even for low energies. This is due to the fact that, as the energy  decreases, the size of the support  of the wavefunction also decreases.  Thus, most  chaotic trajectories evolve in a confined region around the origin and their  close encounters with the  NPXPCs lead to  the same (approximately) limiting distribution. Consequently,  the Born rule is accessible in the long run, only if the initial distribution has the correct ratio between chaotic and ordered trajectories inside the support of the wavefunction.

\item Then we considered coherent states with small amplitudes ($a_0$) for maximum entanglement. In these cases there are also infinite nodal points but they do not come very close to the origin as in the case of large amplitudes. As a consequence the trajectories close to the origin are ordered. Further away we find chaotic trajectories which fill a thick ring around the origin. The distribution of the points in this ring  is irregular for long times but it seems to become uniform after an extremely long time. Therefore these trajectories are probably ergodic in the limit $t\to \infty$.  These trajectories evolve on the outer part of the distribution of the probability density $|\Psi|^2$ and their number is small in the Born distribution, in contrast with the maximally entangled case of large $a_0$ where chaotic-ergodic trajectories dominate the Born distribution and cover all the support of the wavefunction.  We studied in detail the transition from large to small common amplitudes which imply deviations from the two-qubit model. Similar results were found in the case of different amplitudes in $x$ and $y$ directions, where for  large $a_0$ and small $b_0$ we observed that the distribution of the points of the trajectories are concentrated close to the x-axis.

\item Finally  we studied truncated cases with small amplitudes. These cases have both order and chaos and are similar 
to cases studied in previous works of ours \cite{efthymiopoulos2007nodal,contopoulos2008ordered}, where the wavefunction was of the form $\Psi(x,y,t)=\sum_{n_x,n_y}c_{n_x,n_y}Y_{n_x}(x,t)Y_{n_y}(y,t)$. By developing  a general  way to depict graphically the positions of the nodal points of the truncated wavefunction when we do not have their analytical formulae, we found that in the cases with small amplitudes, there are ordered trajectories not only inside the support of the wavefunction (as in the non truncated system) but also at large distances from this support.


\end{enumerate}

In all the aforementioned cases we have considered solutions of Schr\"{o}dinger's equation corresponding to two harmonic oscillators with non-commensurable frequencies. In these cases we find in general  both chaotic and ordered trajectories. 
When the frequencies are commensurable all the trajectories are periodic \cite{tzemos2019bohmian} and therefore there is no chaos at all.


We must emphasize here that all the cases considered above refer to a 2-d system of non-interacting  harmonic oscillators. Thus it would be interesting  to  extend our studies: 

\begin{enumerate}
\item In higher dimensional systems (multipartite systems) {In particular,  we have already shown that in the case of 3 ideal qubits, the increase of the dimensionality of the system implies an increase of the number of the chaotic states for any non zero  entanglement. Thus in multiqubit systems with $n>>3$ we expect that Born's rule is going to be reached practically by any initial distribution of particles} \cite{tzemos2021ergodicity} .
\item In more general Hamiltonians with interaction terms, as e.g. in the case of the quantum H\'{e}non-Heiles model considered in \cite{efthymiopoulos2006chaos}. 
\end{enumerate}

\section*{Acknowledgements}
This research was conducted in the framework of the program of the RCAAM of the Academy of Athens ``Study of the dynamical evolution of the entanglement and coherence in quantum systems.''.

\bibliographystyle{elsarticle-num}
\bibliography{bibliography}

\end{document}